\documentclass{aa}

\usepackage{graphicx}
\usepackage{txfonts}
\usepackage[table,xcdraw]{xcolor}
\usepackage{multirow}
\usepackage[table,xcdraw]{xcolor}

\usepackage{hyperref}
\hypersetup{colorlinks=true, citecolor=blue}

\usepackage{mhchem}
\usepackage{enumitem}

\newcommand{\Msun}{\rm M_{\sun}}
\newcommand{\mstar}{\rm M_{\star}}
\newcommand{\ropt}{\rm r_{\rm opt}}

\newcommand{\mhalo}{\rm M_{\star,\rm halo}}
\newcommand{\mhaloT}{\rm M^{\rm T}_{\star, \rm halo}}
\newcommand{\mhaloLow}{\rm M_{\star, \rm halo, low}}
\newcommand{\mhaloHigh}{\rm M_{\star, \rm halo, high}}
\newcommand{\mhaloInt}{\rm M_{\star, \rm halo, int}}

\newcommand{\mvir}{\rm M_{\rm vir}}
\newcommand{\mgal}{\rm M_{\star,\rm gal}}
\newcommand{\rvir}{\rm r_{\rm vir}}
\newcommand{\minfall}{\rm M^{\rm sat}_{\star,\rm infall}}
\newcommand{\zinfall}{\rm z^{\rm sat}_{{\rm infall}}}

\newcommand{\zdist}{z_{\rm disr}^{\rm sat}}
\newcommand{\agemed}{\rm age_{50\%}}
\newcommand{\cielo}{{\small CIELO}}
\newcommand{\msatmc}{\rm M_{\star, \rm infall}^{\rm sat}}

\begin{document}

   \title{Unveiling the formation channels of stellar halos through their chemical fingerprints}
   
   \author{Jenny Gonzalez-Jara
          \inst{1}\inst{2}\fnmsep\thanks{E-mail: jagonzalez11@uc.cl}
          \and
Patricia B. Tissera\inst{1}\inst{2}\inst{3} \and
Antonela Monachesi\inst{4} \and
Emanuel Sillero\inst{1}\inst{3} \and
Diego Pallero\inst{5} \and
Susana Pedrosa\inst{6} \and
Elisa A. Tau\inst{4} \and
Brian Tapia-Contreras\inst{1}\inst{2} \and
Lucas Bignone\inst{6}
          }
        \institute{Instituto de Astrofísica, Pontificia Universidad Católica de Chile. Av. Vicuña Mackenna 4860, Santiago, Chile.
         \and
Centro de AstroIngeniería, Pontificia Universidad Católica de Chile. Av. Vicuña Mackenna 4860, Santiago, Chile.
        \and 
Núcleo Milenio ERIS, ANID.
        \and
Departamento de Astronomía, Universidad de La Serena, Av. Raúl Bitrán 1305, La Serena, Chile.
        \and
Departamento de Física, Universidad Técnica Federico Santa María, Avenida España 1680, Valparaíso, Chile.
        \and
        Instituto de Astronomía y Física del Espacio, CONICET-UBA, Casilla de Correos 67, Suc. 28, 1428, Buenos Aires, Argentina.
        }

   \date{Received October 10, 2024 ; accepted November 21, 2024}

  \abstract
    {
Stellar halos around galaxies contain key information about their formation and assembly history. Using simulations, we can trace the origins of different stellar populations in these halos, contributing to our understanding of galaxy evolution.}
	{We aim to investigate the assembly of stellar halos and their chemical abundances in 28 galaxies from \cielo~cosmological hydrodynamical zoom-in simulations, spanning a broad range of stellar masses, $\mgal \in 10^9 - 10^{11}\: \Msun$. }
	{
Stellar halos were identified using the AM-E method, focusing on the outer regions between the 1.5 optical radius and the virial radius. We divided the stellar populations based on their formation channel: ex-situ, endo-debris, and in-situ, and analyzed their chemical abundances, ages, and spatial distributions. Additionally, we explored correlations between halo mass, metallicity, and alpha-element enrichment.}
    {\cielo~simulations reveal that stellar halos are predominantly composed of accreted material, including both ex-situ and endo-debris stars, in agreement with previous works. The mass fraction of these populations is independent of stellar halo mass, though their metallicities scale linearly with it. Ex-situ stars tend to dominate the outskirts and be more $\alpha$-rich and older, while endo-debris stars are more prevalent at lower radii and tend to be less $\alpha$-rich and slightly younger. Massive stellar halos ($\mhalo > 10^{9.5} \Msun$) require a median of five additional satellites to build 90\% of their mass, compared to lower-mass halos, which typically need fewer (median of 2.5) and lower-mass satellites and are assembled earlier. The diversity of accreted satellite histories results in well-defined stellar halo mass-metallicity and [$\alpha$/Fe]-[Fe/H] relations, offering a detailed view of the chemical evolution and assembly history of stellar halos. We find that the [$\alpha$/Fe]-[Fe/H] is more sensitive to the characteristics and star formation history of the contributing satellites than the stellar halo mass-metallicity relationship. }
    {}

\maketitle

\section{Introduction}

Stellar halos are one of the visible counterparts of galaxies, that store critical information about their formation histories. Globally, their origin can be explained as a natural consequence of the current cosmological paradigm, Lambda cold dark matter ($\Lambda$CDM).
Within this scenario, dark matter halos are formed through the hierarchical merger of smaller substructures or building blocks \citep{White_1973}. Within dark matter halos, galaxies emerged and evolved, accreting gas and/or satellite galaxies, which contribute to imprinting chemical abundance patterns in the stellar populations.
Due to the long dynamical timescales in halos \citep{Helmi_2000}, the accreted stars from a single progenitor galaxy can be connected through their chemical abundance patterns and integral motions, i.e. energy-angular momentum (E - L$_z$). These chemo-dynamical patterns act as a fossil record of the interstellar medium (ISM) from which stars were born in the contributing satellites \citep[e.g.,][and references therein]{Freeman_2002}. They provide a means to disentangle the history of the assembly of galaxies and also offer insights into the formation of dwarf galaxies in the early Universe \citep[e.g.,][and references therein]{Helmi_2020, Horta_2022, Buder_2022, Mori_2024, Deason_2024}.

The stellar halo of our Galaxy has been continuously scrutinized in the search for the satellite remnants, with several events already been identified \citep[e.g.,][]{Norris_1994, Helmi_2000, Helmi_2008, Carollo_2010, Beers_2012, Conroy_2022}.
Among them, the main accretion event, the so-called Gaia-Enceladus-Sausage (GES) \citep{Helmi_2018, Belokurov_2018}, is a landmark in the history of the formation of the stellar halo of our Galaxy \citep{Gallart_2019, Bignone_2019}. In recent years, a significant contribution to the archaeological study of the Galactic halo has been made by the astrometric mission GAIA \citep{Survey_Gaia_DR1_2016, Survey_Gaia_DR2_2018, Survey_Gaia_DR3_2020} with (future) synergy from others, suchs as LAMOST\citep{Ciuca_2022}, APOGEE \citep{Survey_Majewski_2017}, GALAH \citep{Buder_2022}, 4MOST \citep{Survey_4most}, or DESI \citep{Cooper_2023}.

Extending the study of stellar halos beyond our Galaxy remains challenging due to their low surface brightness characteristics. Nevertheless, by using different observational approaches, several stellar halos have been analyzed in the Local Group and beyond, with a variety of density profiles, kinematics, and metallicity distributions being found \citep[e.g.,][]{Monachesi_2016a, Harmsen_2017, Gilbert_2019, Ogami_2024}. Interestingly, a lack of stellar halos has been claimed in a few galaxies, such as M101 \citep{VanDokkum_2014}, NGC 1042, and NGC 3351 \citep{Merritt_2016}. However, it was later demonstrated using Hubble Space Telescope (HST) observations that M101 actually has a low-mass stellar halo \citep{Jang_2020}. Thus, it is still unclear whether these findings are observational artifacts because of the low surface brightness typical of the stellar halos. Hence, theoretical studies including a wide range of stellar halo masses are crucial to understanding the formation channels across a diversity of galaxies.

Different formation and accretion histories of galaxies combined with different stellar nucleosynthesis channels can produce distinct chemical abundance patterns in galaxies. As a result, a correlation between metallicity and stellar mass is expected \citep{Tinsley_79}. Observationally, \citet{Lequeux_1979} and subsequent studies \citep[e.g.,][]{Tremonti_2004, Lee_2006} provided evidence of the so-called galaxy mass-metallicity relation (MZR). \citet{Harmsen_2017} showed similar behavior for the stellar halos of a set of galaxies observed with HST, particularly using the GHOSTS (Galaxy Halos, Outer disks, Substructure, Thick disks, and Star clusters) survey \citep{GHOST_2011, Monachesi_2016a}. These authors measured the metallicity of halo stars at 30~kpc along the minor axis. This correlation has been attributed to a single accretion event that contributes to the bulk of accreted material and determines its metallicity \citep{DSouza_2018, Monachesi_2019}.
This suggests a strong connection between the chemical properties of the stellar halo and its accretion history, which can be used to infer the masses of the largest accreted satellites \citep{Tissera_2012}.

On the other hand, the ratio of oxygen to iron encodes information about the star formation history of a galaxy due to the interplay between the chemical production of oxygen and iron by the progenitors of type Ia and type II Supernovae (SNIa and SNII, respectively) with different lifetimes. Therefore, [O/Fe]-[Fe/H] is a fundamental plane to understanding the chemical evolution of galaxies \citep{ Matteucci_2021}, because it provides information on the star formation history, initial mass function (IMF), and inflows and outflows events in galaxies \citep[e.g.,][Silva in prep.]{Matteucci_1990, Mason_2023}. The $\alpha$ plane has been used to interpret the origin of different stellar populations in the disk and bulge of the Milky Way in observational and theoretical works \citep{Matteucci_1986, Francois_2004, Zoccali_2006, Hawkins_2015}
The stellar populations of dwarf galaxies occupy distinctive locations on the [O/Fe]-[Fe/H] plane in comparison to those of the Milky Way \citep{Tolstoy_2009, Tissera_2012, Spitoni_2017}.
Modern numerical simulations, which include chemical evolution, have exploited the information stored by the chemo-dynamical patterns to link them to particular events in the history of the evolution of galaxies such as the Milky Way \citep{Mosconi_2001, Belokurov_2020}.

The presence of accreted stars in stellar halos is a natural outcome of $\Lambda$CDM cosmology \citep{Deason_2024}. While their relative dominance varies depending on the specific assembly history of each galaxy, there is broad agreement that the outer regions of stellar halos are predominantly composed of accreted stars (also known as ex-situ stars). These stars could be born in satellites outside the virial radius or formed in infalling gas-rich satellites, when they are already within the virial radius of the main galaxy and are later disrupted \citep[][also called endo-debris]{Tissera_2014}.
This dominant contribution of accreted stars has been reported by numerical simulations, which have mainly focused on the study of Milky Way mass-sized halos \citep{Zolotov_2009, Cooper_2010, Font_2011, Tissera_2012, Monachesi_2019, VeraCasanova_2022, Horta_2023}. Additionally, hydrodynamical simulations also predict the presence of in-situ stars \citep{Zolotov_2009}. Simulations attributed this latter population to different formation channels:
i) stars formed in the disk of the main galaxy and then kicked out to large galactocentric distances due to violent events \citep{Tissera_2014, Monachesi_2019, Belokurov_2020, Khoperskov_2022a},
ii) stars formed from stripped gas in the halo of the main galaxy \citep{Zolotov_2009, Cooper_2015, Monachesi_2019}.

The stellar populations in the halo exhibit particular chemical abundances. \cite{Tissera_2013} studied six Milky Way-like galaxies and identified three stellar halo subpopulations according to their sites of formation, which exhibit different locations on the [O/Fe]-[Fe/H] plane. Ex-situ stars were found to be $\alpha$-enhanced, and have low metallicities, whereas those formed in gas-rich subgalactic systems within the virial radius of the main progenitor galaxies (i.e., wet mergers) had lower [O/Fe]. Disk-heated stars showed low $\alpha$-enhancement and high [Fe/H], as was expected for a population formed from the ISM in disks, which have more time to be enriched by SNIa events. However, there was a large dispersion in the abundances, which reflects the diversity of assembly histories even in systems with similar virial masses. The [O/Fe]-[Fe/H] plane was also used by \citet{Brook_2020} to study the properties of the stars associated with a GES-like event in comparison to stars in the disk of the Milky Way. Hence, most studies have analyzed Milky Way or M31 mass-type galaxies, thereby under-sampling the range of possible stellar halos.

This work aims to investigate the assembly of stellar halos and their chemical abundances in galaxies spanning a broad range of stellar masses, $\mgal \in [10^{9}-10^{11}\Msun]$. We focus on the chemical abundances of the different subpopulations in the stellar halo and link both the MZR and the [O/Fe]-[Fe/H] plane to their formation histories. In this paper, we focus on the outer region of the stellar halos, beyond the central part, which coexists with the main galaxy. In a forthcoming paper, we shall extend this analysis to the inner region of the stellar halo.

For this purpose, we used simulated galaxies from the suite of fully cosmological hydrodynamical zoom-in simulations of Chemo-dynamIcal propertiEs of gaLaxies and the cOsmic web, dubbed \cielo~ \citep{Tissera_2024}. The \cielo~ project aims to study the formation of galaxies in different environments, excluding clusters of galaxies. These simulations have been used in previous studies to investigate the effects of the environment on infalling satellite galaxies \citep{Rodriguez_2022}, the impact of baryons on the shape of dark matter halos \citep{Cataldi_2023}, the shape of the metallicity profiles in galaxies during its evolutionary process \citep{Tapia_2022}, and the feedback effect of the accretion of primordial black holes on gas properties at very higher redshift \citep{Casanueva_2024}.

The paper is organized as follows. In Section \ref{secc:simulations}, we describe the \cielo~ simulations. Section \ref{sec:data} details the sample of galaxies analyzed, and the decomposition method adopted to identify the galactic components. In Subsection \ref{sec:pop}, we define the populations identified in the stellar halos and analyze their different formation channels. In Section \ref{secc:assembly}, we examine the star formation and accretion histories of the \cielo~stellar halos. In Section~\ref{sec:chemical_abund}, we investigate if the chemical abundances might be used as an indicator of the different formation channels through the MZR (in sec \ref{ssec:mzr}), and in \ref{sec:OFe_FeH}, we link the stellar halo enrichment history to their accreted satellites through the [O/Fe]-[Fe/H] plane. Finally, in Section \ref{secc:summary}, we summarize our main results.

\section{CIELO simulations}\label{secc:simulations}

In this work, we analyze stellar halos from six zoom-in simulations of the \cielo~ Project.
The initial conditions correspond to zoom-in regions selected from large cosmological volumes mapping a variety of environments including small groups, voids, walls, and filaments \citep{Tissera_2024}.
All simulated regions were selected from a dark matter-only simulation run of a cosmological simulation consistent with the $\Lambda$CDM universe model with $\Omega_0$ = 0.317, $\Omega_{\Lambda}$ = 0.682, $\Omega_{B}$ = 0.049, $h$= 0.671, $\sigma_{8} = 0.834$, $n_{s}=0.962$ \citep{Planck2014}.

These simulations were run using a version of {\small GADGET-3}, an updated version of {\small GADGET-2} \citep{Springel_2003, Springel_2005}. This version includes a multiphase model for the ISM, metal-dependent radiative cooling, stochastic star formation, and energy and chemical SN feedback, as is described in \citet{Scannapieco_2005, Scannapieco_2006}.
The initial conditions (ICs) were generated at two resolution levels: L12 and L11. For L12, the dark matter particle mass is $\rm m_{\rm dm}= 1.36\times 10^5 \Msun h^{-1}$, while for L11, it is $\rm m_{\rm dm}=1.28\times 10^6 \Msun h^{-1}$, referred to as high resolution and intermediate resolution, respectively. The initial gas masses are $\rm m_{\rm gas}= 2.1\times 10^4 \Msun h^{-1}$ for L12 and $\rm m_{\rm gas}=2.0\times 10^5 \Msun h^{-1}$ for L11. The gravitational softening lengths for simulations at L11 resolution are $\epsilon_{\rm g}= 400$ pc for gas and stellar particles and 800 pc for dark matter particles, while L12 runs have softening lengths of 250 pc for gas and stellar particles and 500 pc for dark matter particles.

The chemical evolution model includes the enrichment by SNIa and SNII, following 12 different chemical isotopes: H,\ce{^{4} He}, \ce{^{12} C}, \ce{^{16} O}, \ce{^{24} Mg}, \ce{^{28} Si}, \ce{^{56} Fe}, \ce{^{14} N}, \ce{^{20} Ne}, \ce{^{32} S}, \ce{^{40} Ca}, and \ce{^{62} Zn} \citep{Mosconi_2001}. The multiphase and SN feedback models have been used to successfully reproduce the star-formation activity of galaxies during quiescent and starburst phases. They are able to trigger violent galactic mass-loaded winds without introducing mass-scale parameters \citep{Scannapieco_2008}.

Virialized structures in \cielo~simulations were identified using a friends-of-friends algorithm \citep[FoF;][]{Davis_1985}. Then, the SUBFIND algorithm \citep{Springel_2001, Dolag_2009} was applied to identify substructures (hereafter, subhalos) within each dark matter halo at all available redshifts. Finally, merger trees were built using the AMIGA algorithm \citep{Knollmann_2009}. We would like to point out that algorithms to identify substructures have difficulties disentangling the contribution of each galaxy during the last stages of mergers, when the systems are very close. Different authors adopt different criteria to make the final decision. Hence, caution should be taken when comparing results from different authors. In the next section, we explain the criteria adopted in this work.

\begin{figure*}
    \centering
    \includegraphics[scale=0.62]{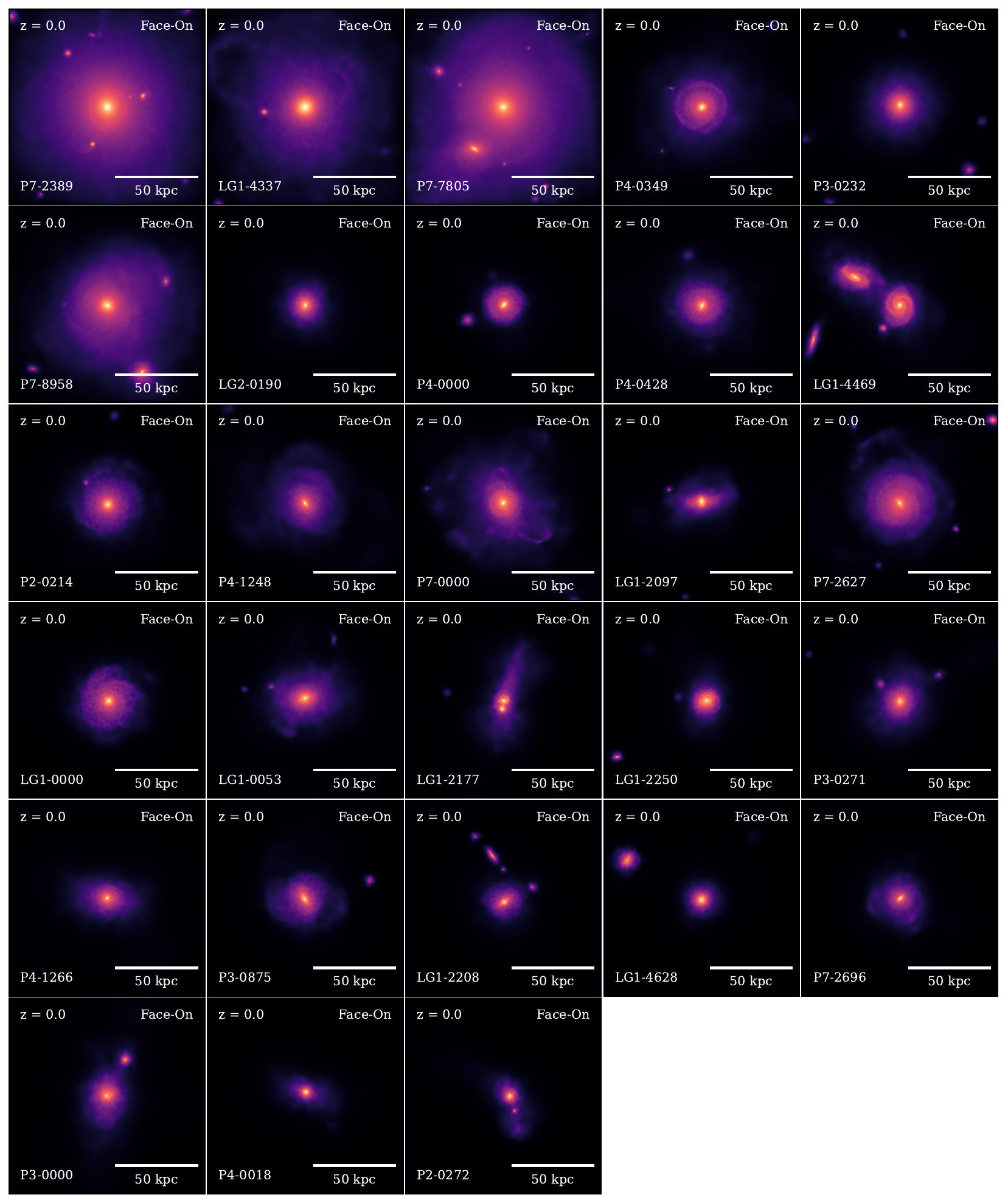}
    \caption{Face-on view of the projected stellar component at $z=0$ for the 28 galaxies analyzed in our sample. Each panel is centered at the center of mass of the corresponding central galaxy and shows all the stellar particles in a cubic volume of 120~kpc size. The panels are ordered by decreasing stellar mass. The \cielo~galaxy ID is displayed in the lower-left corner of each panel.}
    \label{fig:galaxies_faceon}
\end{figure*}

\section{Stellar halos of \cielo~ galaxies}\label{sec:data}

We worked only with central galaxies, which are defined as the most massive Subfind system in each FoF halo.
We selected central galaxies with stellar masses of $\mgal \geq 10^9 \Msun$ without any further restriction. The stellar mass is defined by all stars within $2 \ropt$ of the central galaxies, where $\ropt$ is the radius that enclosed the $83\%$ of the stellar mass of the galaxy.

Our \cielo~galaxy sample reproduces various properties of observed galaxies, including the mass-size relation, and the MZR for both gas and stellar populations \citep{Tissera_2024}. Figure \ref{fig:galaxies_faceon} displays the face-on view of the 28 central galaxies in our sample, highlighting their diverse sizes, masses, and environments.
The simulated galaxies are dynamically decomposed in bulge, disk, and stellar halos, as is explained in the following subsection \ref{ssec:seleccion}.

{\renewcommand{\arraystretch}{1.1}
\begin{table*}[]
\caption{Main properties of the 28 \cielo~ galaxies analyzed.} 
\centering
    \begin{tabular}{c c c c c c c c c}
    \\ \hline \hline
Simulation & ID Gal & $\ropt$ & $\rvir$ & $\mgal$ & $\mhaloT$ & $\mhalo$ & $\mvir$\\  
& & [$\rm kpc$] & [$\rm kpc$] & [$\Msun$] & [$\Msun$] & [$\Msun$] & [$\Msun$]\\ \hline 

LG1 & 4337 & 7.08 & 159.99 & 5.14e+10 & 1.96e+10 & 8.09e+09 & 1.37e+12 \\
LG1 & 4469 & 7.33 & 137.02 & 8.06e+09 & 2.92e+09 & 1.03e+09 & 5.39e+11 \\
LG1 & 2097 & 5.94 & 102.49 & 4.70e+09 & 1.82e+09 & 5.89e+08 & 2.25e+11 \\
LG1 & 0000 & 11.35 & 109.50 & 4.60e+09 & 8.91e+08 & 1.64e+08 & 1.76e+11 \\
LG1 & 0053 & 9.60 & 94.29 & 4.36e+09 & 2.19e+09 & 4.82e+08 & 1.66e+11 \\
LG1 & 2177 & 6.54 & 85.46 & 3.61e+09 & 3.02e+09 & 4.29e+08 & 2.09e+11 \\
LG1 & 2250 & 5.78 & 121.94 & 3.36e+09 & 8.90e+08 & 2.23e+08 & 7.59e+10 \\
LG1 & 2208 & 5.83 & 79.62 & 2.51e+09 & 1.16e+09 & 1.79e+08 & 1.59e+11 \\
LG1 & 4628 & 4.29 & 75.44 & 2.44e+09 & 9.20e+08 & 1.63e+08 & 1.24e+11 \\
LG2 & 0190 & 8.53 & 115.24 & 1.46e+10 & 7.35e+09 & 1.61e+09 & 5.08e+11 \\
P2 & 0214 & 9.57 & 85.67 & 7.22e+09 & 2.80e+09 & 3.83e+08 & 1.99e+11 \\
P2 & 0272 & 5.79 & 61.46 & 1.58e+09 & 6.90e+08 & 2.01e+08 & 7.77e+10 \\
P3 & 0232 & 5.89 & 96.77 & 1.65e+10 & 6.57e+09 & 1.16e+09 & 2.84e+11 \\
P3 & 0271 & 8.89 & 88.11 & 3.29e+09 & 1.70e+09 & 3.51e+08 & 1.19e+11 \\
P3 & 0875 & 9.80 & 75.83 & 2.87e+09 & 1.39e+09 & 2.33e+08 & 1.06e+11 \\
P3 & 0000 & 11.19 & 82.66 & 1.92e+09 & 1.17e+09 & 2.55e+08 & 5.05e+10 \\           
P4 & 0349 & 7.60 & 110.63 & 1.84e+10 & 5.31e+09 & 1.60e+09 & 3.16e+11 \\
P4 & 0000 & 4.77 & 88.18 & 1.32e+10 & 3.96e+09 & 5.05e+08 & 2.14e+11 \\
P4 & 0428 & 7.02 & 98.71 & 1.24e+10 & 2.45e+09 & 7.17e+08 & 2.79e+11 \\
P4 & 1248 & 9.92 & 103.27 & 5.86e+09 & 2.72e+09 & 6.64e+08 & 9.51e+10 \\
P4 & 1266 & 9.19 & 86.79 & 3.12e+09 & 1.73e+09 & 3.38e+08 & 8.93e+10 \\
P4 & 0018 & 4.06 & 69.36 & 1.90e+09 & 8.13e+08 & 2.20e+08 & 9.21e+10 \\
P7 & 2389 & 10.25 & 146.55 & 5.43e+10 & 1.94e+10 & 5.26e+09 & 7.99e+11 \\
P7 & 7805 & 10.93 & 146.54 & 4.09e+10 & 2.10e+10 & 6.79e+09 & 9.71e+11 \\
P7 & 8958 & 7.16 & 91.51 & 1.48e+10 & 4.38e+09 & 1.59e+09 & 2.19e+11 \\
P7 & 0000 & 6.35 & 93.28 & 5.38e+09 & 1.77e+09 & 4.18e+08 & 1.09e+11 \\
P7 & 2627 & 9.43 & 71.73 & 4.69e+09 & 1.41e+09 & 2.49e+08 & 1.19e+11 \\
P7 & 2696 & 2.82 & 58.28 & 2.30e+09 & 1.14e+09 & 2.01e+08 & 4.59e+10 \\

\hline \\
\end{tabular}
\tablefoot{From left to right: Simulation code, ID galaxies, $\ropt$ is the characteristic radius that encloses 83 percent of the stellar mass of the galaxy, $\mgal$ is the galaxy stellar mass up to 2 $\ropt$, $\rvir$ is the virial radius, $\mhaloT$ is the stellar halo mass up to virial radius, $\mhalo$ is the stellar halo mass ranging from $1.5 \ropt$ to $\rvir$, and $\mvir$ is the mass of dark matter within the virial radius.}
\label{table:Galaxies}
\end{table*}}

\subsection{Definition of the stellar halos }\label{ssec:seleccion}

Since the definition of the stellar halo is not straightforward, we first considered the global decomposition presented in \citep{Tissera_2012}, the so-called AM-E method, and defined the disks, bulge, and halo self-consistently. The AM-E method is based on a combination of the angular momentum content of the particles and their binding energy. According to this method, galactic components were decomposed dynamically based on the orbital circularity parameter ($\epsilon = \frac{J_z}{J_{\rm z, max(E)}}$), where $J_{\rm z}$ is the angular momentum component perpendicular to the disk plane and $J_{\rm z, max(E)}$ is the maximum $J_{\rm z}$ over all particles of a given binding energy (E).

The bulge was defined as all the stellar particles more bounded than the minimum energy ($E_{\rm cen}$) of stars with $r \geq 0.55 \ropt$. Therefore, the criterion adapts to the overall size of each galaxy. Although there is some arbitrariness in these criteria, the main features of the components change only slightly with the reference energies. The disk component is defined as the stellar particles with $|\epsilon| \geq 0.5$, $E \geq E_{\rm cen}$ and $r \leq 2r_{\rm \ropt}$. Stellar particles that do not belong to the bulge and disk components are considered part of the stellar halo.

The decomposition was performed to all the central galaxies satisfying the mentioned criteria at $z = 0$ and to their progenitors up to $z\approx 6$.
This implies decomposing and analyzing 28 halos across 86 snapshots separated by a median of $1.4 \times 10^{8}$~yrs (with a minimum of $3 \times 10^{7}$~yrs and a maximum of $1.7 \times 10^{8}$~yrs). The cadence of CIELO simulations is appropriated to follow satellite galaxies falling into their main halos and to study the impact that the accreted satellites have on different physical processes, such as the assembly of the stellar halo and the other galactic components.

\begin{figure*}
    \centering
    \includegraphics[scale=1.1]{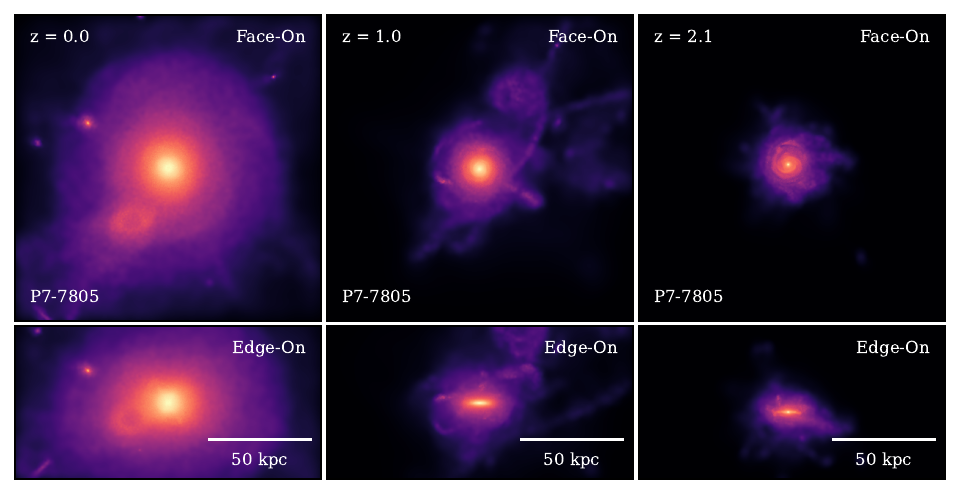}
    \caption{Face-on (upper panel) and edge-on (lower-panel) view of the stellar halo density maps for a single \cielo~ galaxy. Three redshifts are depicted from left to right: 0.0, 1.0, and 2.1. Each panel shows all the stellar particles identified as part of the stellar halo of galaxy P7-7805 (see sec\ref{ssec:seleccion} for details on identifying the galactic components). }
    \label{fig:evol_sh}
\end{figure*}

Due to the difficulty of isolating halo stars in the very inner region of nearby galaxies, or even our own Galaxy, studies of stellar halos with masses similar to the Milky Way typically focus on regions beyond a galactocentric distance of 10~kpc \citep{Harmsen_2017, Monachesi_2019}. We followed a similar criterion, however, considering the wide diversity in galaxy sizes in our sample, we set the galactocentric limit at $1.5\ropt$. Therefore, in this work, we defined the inner halo as the region inside $1.5\ropt$, while the outer region extends within ${1.5\ropt-\rvir}$. For the sake of simplicity, hereafter the outer region will be referred to as the stellar halo, with a stellar mass of $\mhalo$, while the total stellar halo mass will be referred to as $\mhaloT$. The inner halo will be analyzed in a forthcoming work. The separation of the stellar halo in these two regions was performed regardless of the origin of the stellar particles.
We ensured that each stellar halo was sufficiently resolved numerically by setting a lower limit of $\mhalo > 10^8 \Msun$ (equivalent to a minimum number of $\sim 1300$ and a maximum of $\sim 400000$ stellar particles).
Table~\ref{table:Galaxies} summarizes the main properties of the studied galaxies and their stellar halos. In this Table, we also include the \cielo~code for the zoom-in regions: "LG" for Local Group analogue and the "P" for the Pehuen halos\footnote{Pehuen is a term from Mapudungún used to refer to the Araucaria araucana tree, specifically found in Chile and Argentina.}.
In order to analyze the formation of stellar halos, we divided our sample into three $\mhalo$ intervals. Each mass bin is separated by one order of magnitude in mass, i.e. low mass ($\mhalo \in [10^{8.0} - 10^{8.5}]~\Msun$), intermediate ($\mhalo \in [10^{8.5} - 10^{9.5}]~\Msun$) and high mass ($\mhalo \in [10^{9.5} - 10^{10}]~\Msun$).

For illustration purposes, Fig.~\ref{fig:evol_sh} shows the evolution of the stellar halo of a typical galaxy. From left to right, the redshift displays are $z = 0.0$, 1.0, and 2.1. Their viral radius, $\rvir$, varies from to 117~kpc at $z=2.1$ to $\rvir~=147$~kpc at $z=0$. At $z =0.0$ the halo is very smoothed but still shows signs of a later accretion and surviving satellites, which are close to the central region.

At $z = 1.0$ and $z = 2.1$ the stellar halo exhibits clear signals of satellite accretions at different stages of evolution. Well-defined stellar trails along the orbits can be appreciated from this figure. Additionally, the SUBFIND algorithm does not identify the substructures when they become too fluffy and are very close to the central galaxy. This illustrates the complexity of studying stellar halos due to the different components that make them.

\begin{figure}
    \centering
    \includegraphics[width=0.45\textwidth]{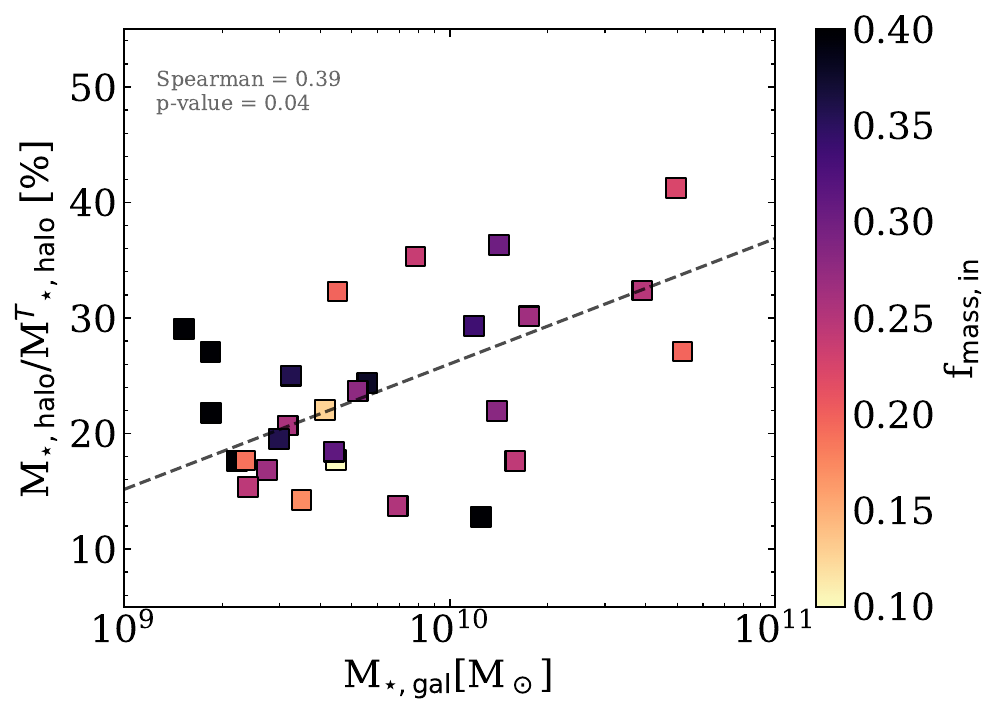}
    \caption{Fraction of the stellar halo mass, $\mhalo$, defined within the range $[1.5\ropt,\rvir]$, with respect to the total stellar halo mass, $\mhaloT$, as a function of the stellar galaxy mass, $\mgal$. Squares are color-coded by the mass fraction of in-situ stars in the stellar halo.}
    \label{fig:Mh_Mgal}
\end{figure}

Figure~\ref{fig:Mh_Mgal} shows the ratio of $\mhalo$ and $\mhaloT$ as a function of the stellar galaxy mass. We can consider this ratio as a proxy of the concentration of a stellar halo. As can be seen, all galaxies have $\sim 10- 40\%$ of their stellar halo mass within ${1.5\ropt-\rvir}$. Therefore, most of the stellar halo mass is concentrated within the inner $1.5\ropt$ (displayed in colors). We obtained a trend so that more massive galaxies have less concentrated stellar halos (Spearman coefficient of $\rm r = 0.36$ and $\rm p = 0.04$). However, at a fixed stellar galaxy mass, there is a diversity in the concentration of their stellar halo, which can be explained due to their different formation histories.
The color code displays the total fraction of in-situ stars, f$_{\rm mass, in}$, identified within the $\rvir$ (see detailed definition in the next section \ref{sec:pop}). We can appreciate a large diversity of f$_{\rm mass, in}$, with a weak trend of halos with larger fractions being associated with less massive galaxies. Since most of the in-situ stars are located within the inner region of the halos, this fact could explain the detected trend. We shall analyze the inner region of the halos in a forthcoming paper, as was mentioned above.

\begin{figure*}
    \centering
    \includegraphics[width=1\textwidth]{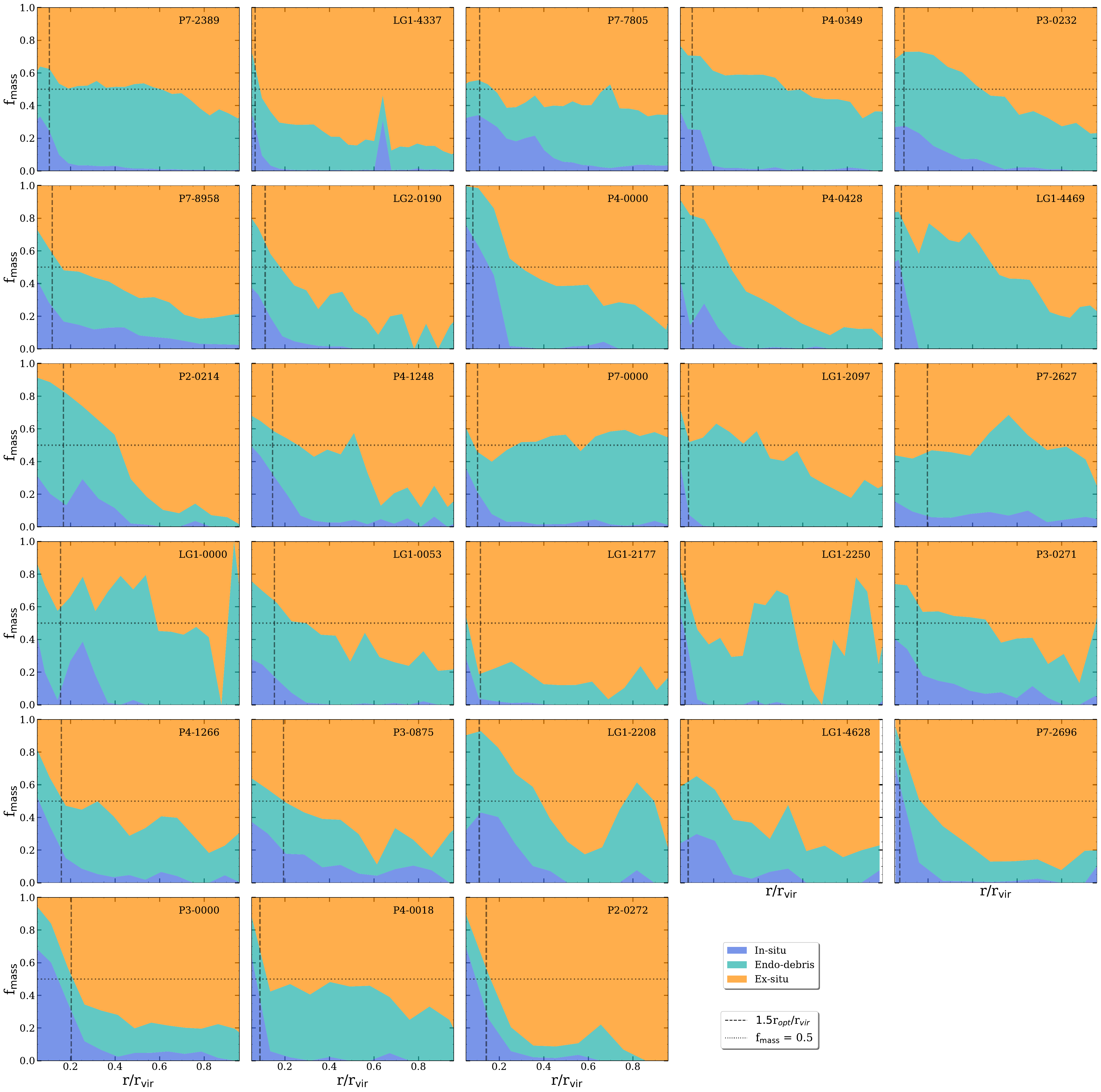}
    \caption{Stellar mass fraction, $f_{\rm mass}$, of each population in the stellar halo: in-situ (blue area), endo-debris (light green area), and ex-situ (orange area) stars as a function of the galactocentric radius in units of its corresponding virial radius, $\rvir$. Stellar halos are sorted by decreasing their stellar galaxy mass, $\mgal$. The vertical line depicts the $1.5\ropt$ in units of $\rvir$. The dashed, horizontal line indicates $f_{\rm mass} =0.5$.}
    \label{fig:pop_vs_r}
\end{figure*}

\subsection{Stellar populations in the stellar halo }\label{sec:pop}

The classification of stellar populations according to their formation channels varies slightly from author to author \citep[e.g.,][]{Norris_1994, Zolotov_2009, Carollo_2010, Tissera_2012, Beers_2012, Pillepich_2014, Buder_2022}. Here, we explain in detail the definitions we adopt to describe and quantify the assembly history of our stellar halos.

In order to identify the origin of each stellar population that built the simulated halos at $z = 0$, we followed them back in time to the closest available snapshot to their birth time and we identified their formation site. According to that, stellar populations are grouped into three types: in-situ, endo-debris, and ex-situ.
 
    \subsubsection{In-situ stars:} Stellar particles gravitationally bound to the central galaxy at the first available snapshot after their birth time.

We identified two channels of formation of in-situ halo stars in \cielo~simulations: disk-heated stars and clumpy in-situ stars.

Here, we defined disk-heated stars as those formed within the galaxy on orbits that initially (at birth-time) confined them to the disk component and became heated to eccentric orbits possibly by mergers and interactions events \citep{Zolotov_2009, Purcell_2010, Tissera_2013}.
The time of formation of a stable disk component varies and, at very high redshift, it could be challenging to define it because of the higher frequency of mergers and interactions \citep{Kassin_2012}. Additionally, numerical resolution in simulations could prevent a clear determination of a disk structure \citep[e.g.,][]{Sestito_2021,Carollo_2023}.
This might lead to fuzzy boundaries between the galactic components, i.e. the bulge, disk, and stellar halo at high redshift.

As we show in Fig.~\ref{fig:evol_sh}, mergers and interactions are ubiquitous and they can also disturb the central galaxy, dynamically heating up stars \citep[e.g.][]{Bignone_2019,Font_2020, Khoperskov_2022a}. Therefore, the contribution of disk-heated stars to the in-situ population depends on the formation history of each galaxy. While these stars are mainly confined to the inner region of the halos, there is a fraction of them in the external ones. These disk-heated stars exhibit a median age of 7.73~Gyr (with a 25-75$^{\rm th}$ at 6.23~Gyr and 9.83~Gyr, respectively) and median [Fe/H] of -0.98~dex, (with a 25-75$^{\rm th}$ at -1.21~dex and -0.80~dex, respectively).

Additionally, we define a category referred to as clumpy in-situ stars.
In this category, we include stars formed at high redshift when the formation is more violent and involves the accretion of several clumps.
These stars are older and metal-poor than disk-heated ones, with a median age of 10.60~Gyr (with a 25-75$^{\rm th}$ at 9.54~Gyr and 11.36~Gyr) and median [Fe/H] of -1.41~dex (with a 25-75$^{\rm th}$ at -1.61~dex and -1.21~dex, respectively).
We also include stars that are associated with disrupted satellites, which have not long identified by the SUBFIND but still have a remnant core. They represent a small fraction of the in-situ population, in some cases less than $1\%$ of the stellar halo. They have a median age of $12^{12.44}_{11.77}\rm{Gyr}$ and a median metallicity of $-2.40^{-2.79}_{-2.00}~\rm{dex}$. There is a negligible fraction (less than $0.1\%$ of in-situ stellar mass) of stars older (median age 13.24~Gyr), for which we were unable to calculate their birth radii. We decided to include them as part of clumpy in-situ halo stars.

    \subsubsection{Endo-debris stars:}Stellar particles formed from gas bound to a subhalo inside $\rvir$, subsequently stripped as they orbit the potential well of the main galaxy.

We traced the history of the gas particles that were detected as stars within each stellar halo at $z=0$, and determined whether they were initially bound to a satellite or to the central galaxy before being transformed into stars. Using this approach, we identify endo-debris stars as those formed in association with the infalling satellites. The main pathway is the formation within the satellites, but a fraction could be born in material that is being disrupted.
Approximately $77^{\scriptstyle{84}}_{\scriptstyle{66}}\%$ of endo-debris stars originate from the first pathway. They have a median age of $10.47^{11.49}_{8.49}$~Gyr, with a median [Fe/H] of $-1.14^{-0.97}_{-1.27}$~dex. The remaining $\sim 23^{\scriptstyle{34}}_{\scriptstyle{16}}\%$ of stars formed in disrupted remnants, during close encounters with other satellites or the central galaxy, while they are still dense and cold. They have median ages of $9.02^{10.26}_{7.25}$~Gyr and median [Fe/H] equal to $-1.16^{-0.99}_{-1.37}$~dex. Since the median differences in their metallicities and ages are less than 0.1 dex and $\sim 1.5$ Gyr, respectively, distinguishing these two formation mechanisms remains debatable. They both come from gas associated with a satellite system and hence we consider them all endo-debris stellar populations.

We highlight that other authors consider endo-debris stars either as part of the in-situ population (stars formed from the stripped gas) or as part of the ex-situ population (stars formed inside of the orbiting satellites that are tidally disrupted and subsequently accreted). However, in this work, the endo-debris population is analyzed independently of in-situ and ex-situ populations because they exhibit different chemical properties that might help link observations with galaxy formation models.

\subsubsection{Ex-situ stars:} Stellar particles gravitationally bound to a subhalo at their birth time but born outside the virial halo of the central galaxy.

Ex-situ stars in \cielo~stellar halos are older than the other stellar halo populations, exhibiting a median age of 11.60~Gyr (with a 25-75$^{\rm th}$ at 10.87~Gyr and 12.33~Gyr, respectively) and a median [Fe/H] of -1.38~dex (with a 25-75$^{\rm th}$ at -1.53~dex and -1.20~dex, respectively). They are formed in other galaxies before they fall into the potential well of our selected central galaxies. Throughout the paper, we use the term 'accreted' material to refer to the ex-situ and endo-debris stars, as part of a large category to describe that they came from either accreted star or gas.

Figure~\ref{fig:pop_vs_r} displays the stellar mass distribution of each population as a function of the galactocentric radius, normalized by the corresponding virial radius. The stellar mass of each population is normalized to the stellar mass in each radial bin. In nearly all galaxies, beyond $1.5\ropt$ the accreted material (ex-situ and endo-debris stars together) is the dominant component, accounting for over $80\%$ of the stellar halo mass in each bin across all radii. While some galaxies exhibit a contribution of in-situ stars at larger radii, their presence comprises fewer than 100 stellar particles per bin in most of the cases, which represents less than $1\%$ of the stellar halo mass. They are stars associated with small satellites that the subfind cannot resolve them (for example, the bump present in LG1-4337 at a large radius in Fig.~\ref{fig:pop_vs_r}).

\begin{figure}
    \centering
    \includegraphics[width=0.45\textwidth]{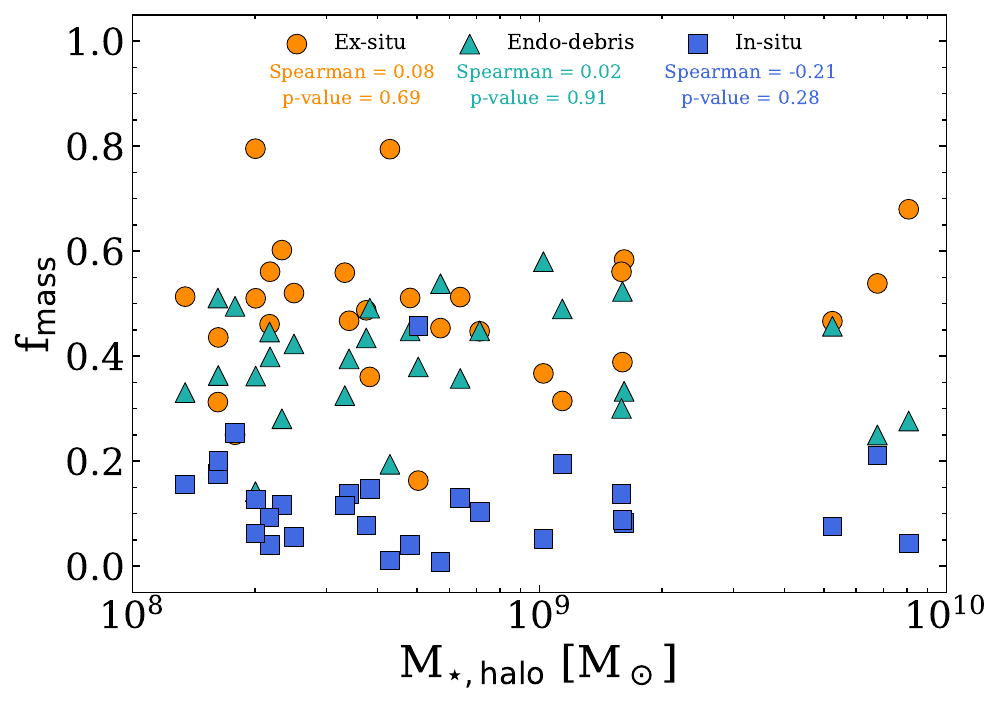}
    \includegraphics[width=0.45\textwidth]{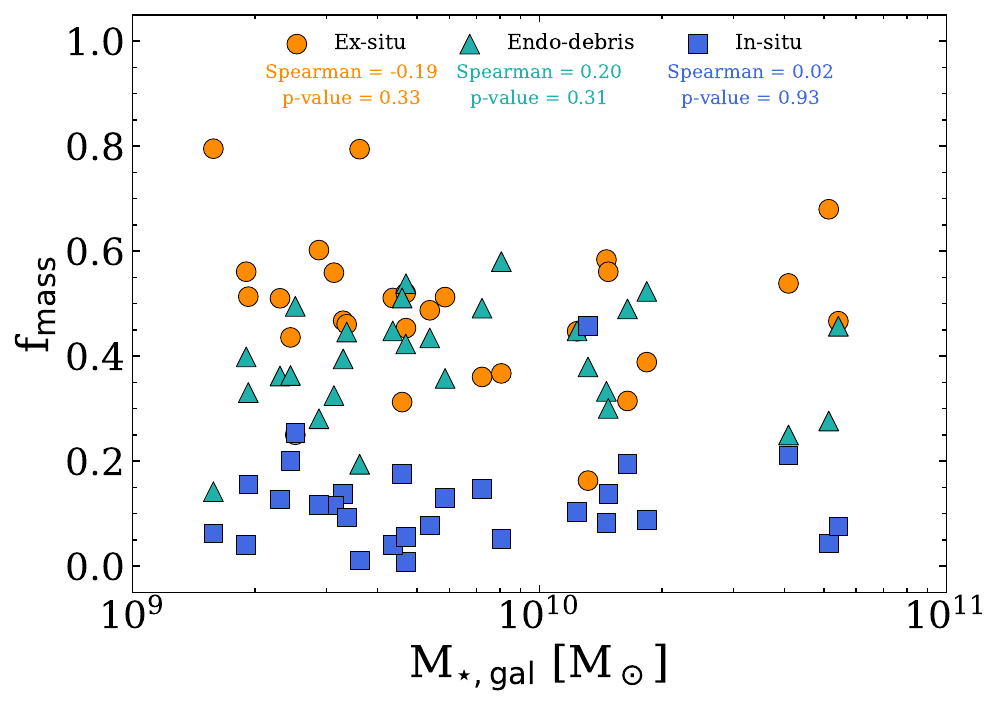}
    \caption{Mass fraction of ex-situ stars (orange circles), endo-debris (light green triangles), and in-situ (blue squares) stellar populations as a function of $\mhalo$ (upper panel) and $\mgal$ (lower panel). We provide Spearman coefficients and p-values in each panel. In the upper panel, we fit linear regressions and represent the standard deviations with shaded areas for each population.}
    \label{fig:Pop_z0}
\end{figure}

Figure~\ref{fig:Pop_z0} displays the mass fraction contributed by the different stellar populations identified in this work as a function of the stellar halo mass (upper panel) and the stellar galaxy mass (lower panel).
Our analysis reveals that the mass fraction of each stellar population, in-situ, endo-debris, and ex-situ, varies from galaxy to galaxy depending on their formation history. However, the Spearman correlation factors indicate no significant dependence on either $\mhalo$ or $\mgal$ as can be seen from Fig.~\ref{fig:Pop_z0}.

Interestingly, Fig.~\ref{fig:Pop_z0} shows that even stellar halos smaller than $10^{9}\Msun$ are composed of the three defined stellar populations, although \cielo~stellar halos are predominantly constituted by accreted material, which makes up to roughly more than $80\%$ of the stellar mass in most of our \cielo~galaxies.
The endo-debris stars and ex-situ stars contribute approximately $40\%$ and $50\%$, respectively, of the stellar material in the stellar halo. On the other hand, in-situ stars contribute in median value $\sim 10\%$ of the stellar halo mass. It is worth noting that our analysis focuses on the region ranging from $1.5\ropt$ to $\rvir$.

\section{Stellar halo assembly}\label{secc:assembly}

Galaxies grow their stellar mass through two mechanisms: forming new stars, which constitute the in-situ population, and accreting satellite galaxies which bring their gas and stars, contributing to the accreted stars category (i.e., ex-situ and endo-debris stars). To gain further insight into these two channels, we analyzed the star formation history (SFHs, Section~\ref{ssec:sfh}) and the accretion history of the \cielo~stellar halos (Section~\ref{ssec:acc}).

\subsection{Star formation history}\label{ssec:sfh}

\begin{figure*}
    \centering
    \includegraphics[width=1\textwidth]{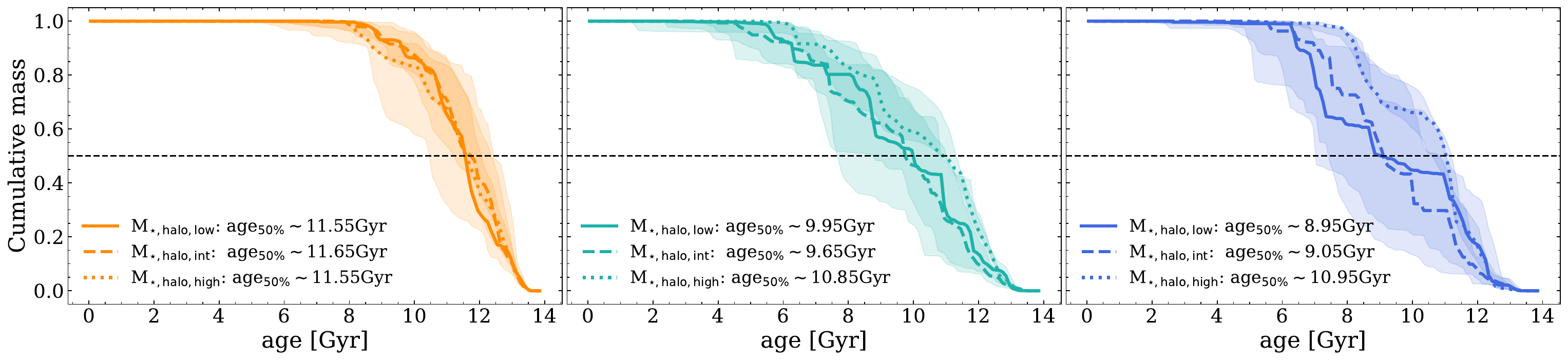}
    \caption{Cumulative stellar mass fraction as
a function of the stellar ages of the three defined stellar populations: in-situ (right panel), endo-debris (middle panel), and ex-situ (left panel). For each stellar halo mass, the legend indicates the median $\agemed$, representing the time when $50\%$ of the final stellar mass was formed. The three $\mhalo$ intervals are shown with different line styles.}
    \label{fig:SFH}
\end{figure*}

We compute the SFH as the cumulative mass per age bin of all stars belonging to the stellar halo at $z=0$. Figure~\ref{fig:SFH} displays the median SFH for ex-situ (left panel), endo-debris (middle panel), and in-situ (right panel) stellar populations for each defined stellar halo mass interval (as is described in Section 3.1). The shaded regions in each panel indicate the $25^{\rm th}$ and $75^{\rm th}$ percentiles of the mass fractions in each age bin. We also include the median age at which $50\%$ of the final stellar mass was formed, $\agemed$.

The left panel of Fig.~\ref{fig:SFH} shows that ex-situ stars are the oldest population within stellar halos. The median $\agemed$ ranges from 11.55~Gyr in less massive halos to 11.65~Gyr in the intermediate ones. The middle panel shows endo-debris stars formed later and their median ages range from 9.65~Gyr for intermediate mass halos to 10.85~Gyr for the more massive ones. Their age distribution is similar regardless of the stellar halo mass. Finally, in-situ stars are, in median values, the youngest population. They show an increase in age while increasing halo mass by 2~Gyr. In-situ stars are composed mainly of disk-heated stars which tend to be young due to the fact that the disks tend to be more actively forming stars at lower redshifts. The youngest ages of in-situ stars in low-mass stellar halos suggest that they have more recent higher star formation activity than massive halos.

In summary, we found a clear distinction in the $\agemed$ and the star formation histories of the selected three stellar halo populations (in-situ, endo-debris, and ex-situ) that composed the simulated stellar halos. The history of the formation of ex-situ and endo-debris stars does not differ significantly from the low to high-mass stellar halo. In the next section, we shall explore the properties of satellites that contribute to the accreted material in the stellar halos.

\subsection{Accretion history}\label{ssec:acc}

As was previously mentioned, stellar halos are predominantly composed of accreted material (i.e., ex-situ and endo-debris stars). When these stars are incorporated into the stellar halos, they imprint distinct chemical abundance patterns. Therefore, to complete our analysis so that we can analyze the chemical patterns in the last Section, it is crucial to explore the assembly histories of our halos in relation to their primary contributing satellites.

For this purpose, all satellites were followed back in time using the merger trees. It is worth mentioning that we adopted a conservative lower limit for halo mass\footnote{We considered only satellites with more than 100 stellar particles at the time they enter the virial radius of the central galaxy, corresponding to a stellar mass limit of $\sim 10^6~\Msun$ and $\sim 10^7~\Msun$ for the resolution levels L12 and L11, respectively (see Section~\ref{secc:simulations}). These stellar mass limits imply that our minimum dark matter halos have $\sim2\times10^7 \Msun$.}, determined by the numerical resolution of our simulations. Lower mass halos down to $10^6~\Msun$ might be able to retain some gas and to contribute with stars with particular chemical signatures to stellar halos \citep[e.g][]{Bland-Hawthorn_2015}. However, exploring such low-mass ranges would require ultra-higher-resolution simulations and their contribution would be too small, so our results would not be significantly impacted.

To improve clarity, we define below key concepts that will be referenced throughout this section:

\begin{itemize}

    \item Infall redshift ($\zinfall$): Redshift of the last snapshot before the satellite enters the virial radius of the central galaxy. If a satellite enters more than one time, then we consider the first time it does so.\\
    
    \item Stellar halo main contributor (SHMC): Satellite that contributes with the highest amount of accreted stellar mass (both ex-situ and endo-debris together) to the stellar halo of a central galaxy.\\
    
    \item Stellar mass of SHMC ($\minfall$): Stellar mass of the main contributing satellite at $z_{\rm infall}$. All the stellar particles within $1.5\ropt$ of the satellite are considered.\\

    \item Accreted stellar mass (M$_{\rm acc}$): Total mass of stellar particles born in accreted satellite galaxies either before (i.e., ex-situ) or after (i.e., endo-debris) $\zinfall$.\\
    
    \item Surviving satellites: Accreted satellites that survived as separated subhalos within $\rvir$ of a central galaxy at $z = 0$. They have contributed with stars or gas to the stellar halo of the central galaxy.\\

    \item Destroyed satellites: Accreted satellites disrupted after $\zinfall$ within the stellar halo of the central galaxy.\\
    
    \item Disruption redshift ($\zdist$): Redshift at which the accreted satellite is no longer recognized as a separate structure and is considered bound to the halo of the central galaxy. This time was estimated by using the merger tree.

\end{itemize}

\begin{table}[]
\begin{center}
\caption{Median number of satellites for each stellar halo mass interval.}
\renewcommand{\arraystretch}{1.5} 
\begin{tabular}{cccc}
\cline{2-4}
& \multicolumn{3}{c}{Number of satellite} \\ \cline{2-4} 
\multicolumn{1}{c}{} & \multicolumn{1}{c}{$\mhaloLow$} & \multicolumn{1}{l}{$\mhaloInt$} & \multicolumn{1}{c}{$\mhaloHigh$} \\ \hline
\multicolumn{1}{c}{95$\%$M$_{acc}$} & $2.5^{3}_{2}$ & $5^{7}_{4}$ & $11^{13.5}_{11}$ \\
\multicolumn{1}{c}{90$\%$M$_{acc}$} & $2.5^{3}_{2}$ & $5^{5.5}_{3}$ & $8^{9}_{8}$ \\
\multicolumn{1}{c}{50$\%$M$_{acc}$} & $1^{1}_{1}$ & $1^{2}_{1}$ & $2^{2}_{2}$ \\ \hline
\end{tabular}
\end{center}
\tablefoot{Median number of satellites needed to reach the $50\%$, $90\%$ and $95\%$ of the accreted stellar halo material. Each column represents a stellar halo mass interval, i.e. from the left to right side: low, intermediate, and high stellar halo mass. }
\label{table:sat}
\end{table}

\begin{figure*}
    \centering
    \includegraphics[width=1\textwidth]{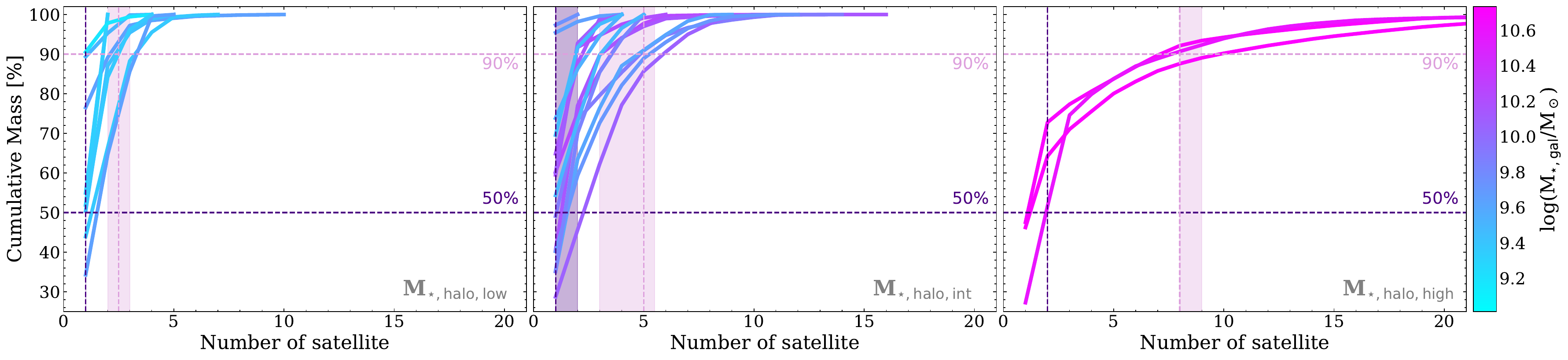}
    \caption{Cumulative accreted stellar mass per number of satellites order according to the relevance of their contribution. Each panel represents a stellar halo mass interval (according to the corresponding titles). The vertical lines denote the median number of satellites required to reach the $90\%$ (dashed pink lines) and $50\%$ (dashed-dotted violet line) of the stellar accreted halo mass. The shades regions denote the 25-75$^{\rm th}$ of satellites needed to reach the $90\%$ (pink shades) and $50\%$ (violet shades).}
    \label{fig:Nsat}
\end{figure*}

Figure~\ref{fig:Nsat} shows the percentage of cumulative accreted stellar mass contributed per satellite, either disrupted or surviving, to the stellar halos at $z=0$. The horizontal lines depict the $95\%$ and $50\%$ of the accreted stellar mass. We estimated the median number of satellites required to reach 50 and 90 percent of the accreted stellar mass in each $\mhalo$ interval. The median number of satellites required to reach the $90\%$($50\%$) of M$_{\rm acc}$ ranges from 2.5(1) to 8(2) from low-mass to high-mass halos (see also Tau et al. in prep. for a similar analysis with the Auriga low-mass simulated galaxies). In Table~\ref{table:sat}, we summarize the median numbers per stellar halo mass and the 25-75$^{\rm th}$ percentiles.

\cite{Monachesi_2019} analyzed 28 Milky-Way mass-liked galaxies from the Auriga simulations and found that the number of satellites that contribute $90\%$ of the accreted stellar mass varies from 1 to 14 with a median of 6.5. We note that these authors considered all accreted stars within $\rvir$ but outside 5 kpc from the galactic center. Their stellar halo masses range from $\sim 7\times10^{9}$ to $5\times10^{10}$ M$_\odot$, corresponding to our high-mass interval. Our findings are consistent with this mass range. Stellar halos more massive than $10^{9.5} \Msun$ require, on a median, five more satellite galaxies to reach $90\%$ of the stellar halo mass compared to low-mass stellar halos.
In general, \cielo~stellar halos with $\mhalo<10^{9.5}\Msun$ need a median of one and four satellites to reach $50\%$ and $90\%$ of their stellar mass, respectively. Hence their SHMCs play a more significant role in determining the properties of the stellar halos.

\begin{figure}
    \centering
    \includegraphics[width=0.45\textwidth]{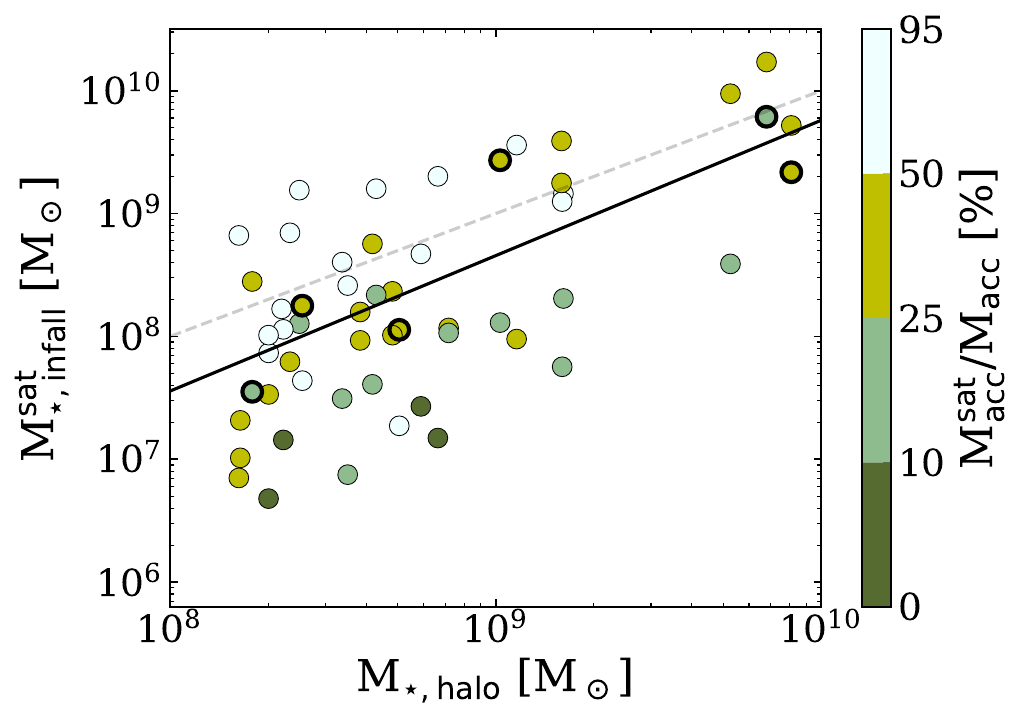}
    \includegraphics[width=0.45\textwidth]{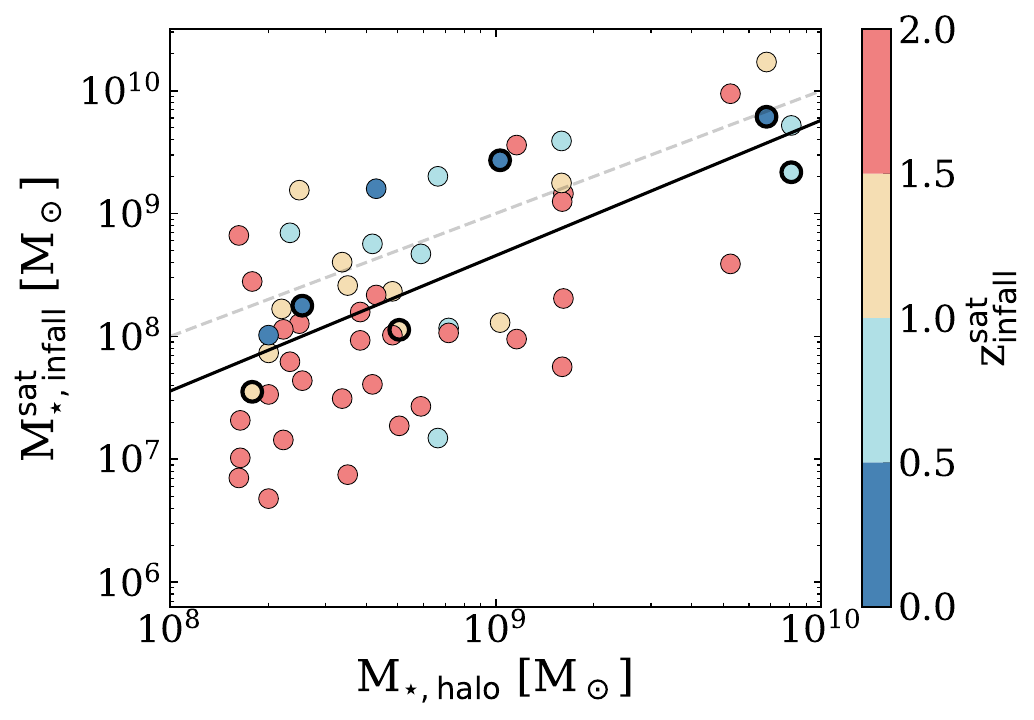}
    \includegraphics[width=0.45\textwidth]{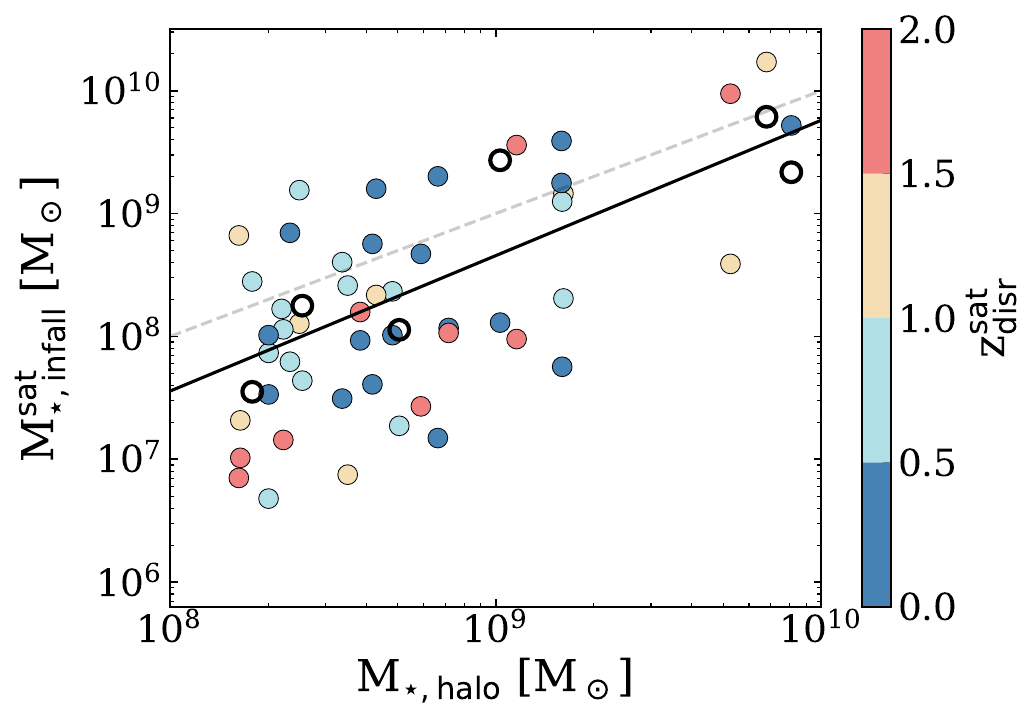}
    \caption{Stellar mass ($\minfall$) of the two main contributor satellites (SHMCs) to the stellar halo as a function of the stellar halo mass. The upper panel is color-coded by the percentage mass contributed by the SHMC to the accreted stellar halo (M$_{acc}$), the middle panel is color-coded by the infall redshift of the satellite ($\zinfall$), and the lower panel is color-coded by the disruption redshift ($\zdist$). In some galaxies, the main contributor is a surviving satellite which is denoted by open circles. The dashed line represents the 1:1 line, while the solid line indicates the linear regression.}
    \label{fig:sats_caract}
\end{figure}

Following previous works, which reported that Milky Way mass-sized halos have mostly one or two main contributions that bring in most of the accreted material \citep{DSouza_2018, Monachesi_2019, Fattahi_2020}, and in order to gain insight into the assembly histories and the properties of the stellar halos, we analyze the two main satellite contributors (SHMCs). Figure~\ref{fig:sats_caract} shows $\minfall$ of the two SHMCs as a function of the stellar halo mass, color-coded by the percentage mass contribution relative to total M$_{\rm acc}$ (upper panel), of $\zinfall$ (middle panel), and of $\zdist$ (lower panel). There is a correlation (Spearman coefficient = 0.56, p-value $\sim 0$) between $\minfall$ and $\mhalo$. This trend is consistent with the previous results by \citet{Bell_2017}.
The solid line represents a linear fit which provides a rough prediction of the stellar mass of the SHMCs of the stellar halos within our covered dynamical range: $\rm{log}_{10}\minfall$ = $1.10\times \rm{log}_{10}\mhalo -1.26$.

The upper panel of Fig.~\ref{fig:sats_caract} illustrates that a single SHMC contributes significantly more to low-mass halos as shown before, accounting for over $50\%$ of the accreted stellar halo in some cases. In contrast, massive stellar halos ($\mhalo > 10^{9.5} \Msun$) tend to have SHMCs that are more massive than those in low-mass stellar halos, yet they contribute smaller fractions to the overall stellar halo because they tend to sink further into the potential well reaching the central regions, before being completely disrupted. This explains the fact that some of the satellites lie above the 1:1 line. This occurs because the $\msatmc$ represents an upper mass limit, as certain stars end up in the central regions of the galaxy rather than outside $1.5\ropt$ where the stellar halo is defined.

The middle panel of Fig.~\ref{fig:sats_caract} shows a diversity of $\zinfall$, with a trend for more massive halos to have more recent accretion of satellites as was expected. Conversely, low-mass halos acquired most of their SMHCs before $z \sim 1.5$ (media lookback time of $\sim 10$ Gyr compared to $\sim 8$ Gyr in agreement with results from Tau et al. in prep. using the Auriga simulations). However, due to the effects of dynamic friction, the disruption of these satellites takes longer than more massive satellites, as can be seen in the lower panel of this figure. Some galaxies have a SHMC that can still be identified at $z=0$, as a surviving satellite (denoted by open circles in the middle and lower panel of Fig.~\ref{fig:sats_caract}). As can be seen, we detected stellar halos for which neither of the SHMCs contributed with more than 50 percent of $\mhalo$.

\begin{figure}
    \includegraphics[width=0.45\textwidth]{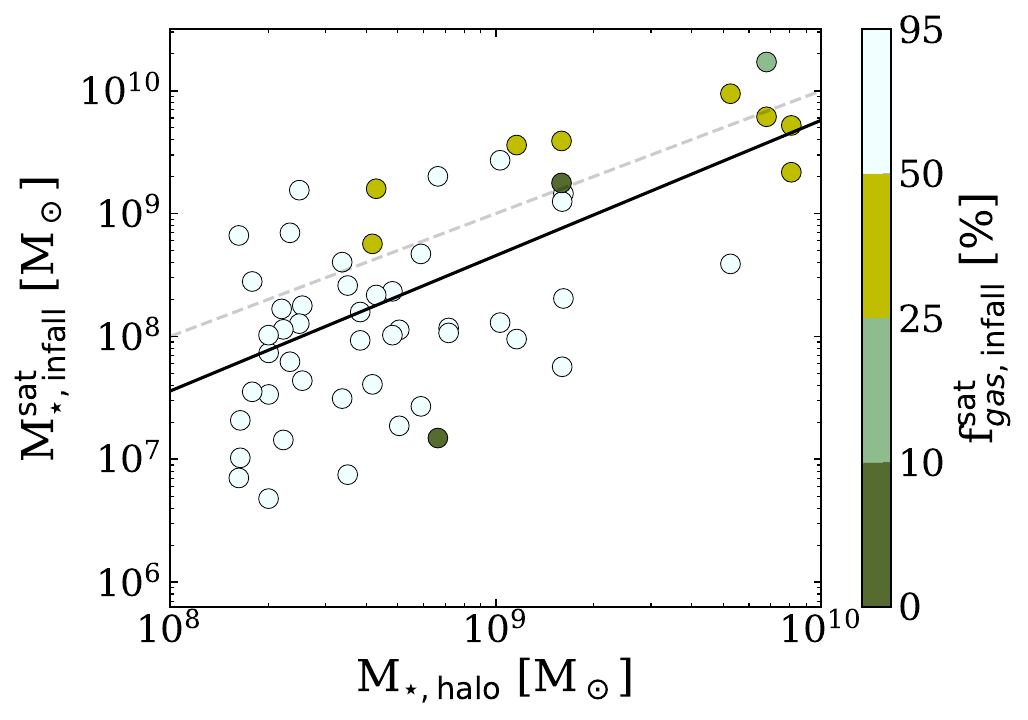} 
    \includegraphics[width=0.45\textwidth]{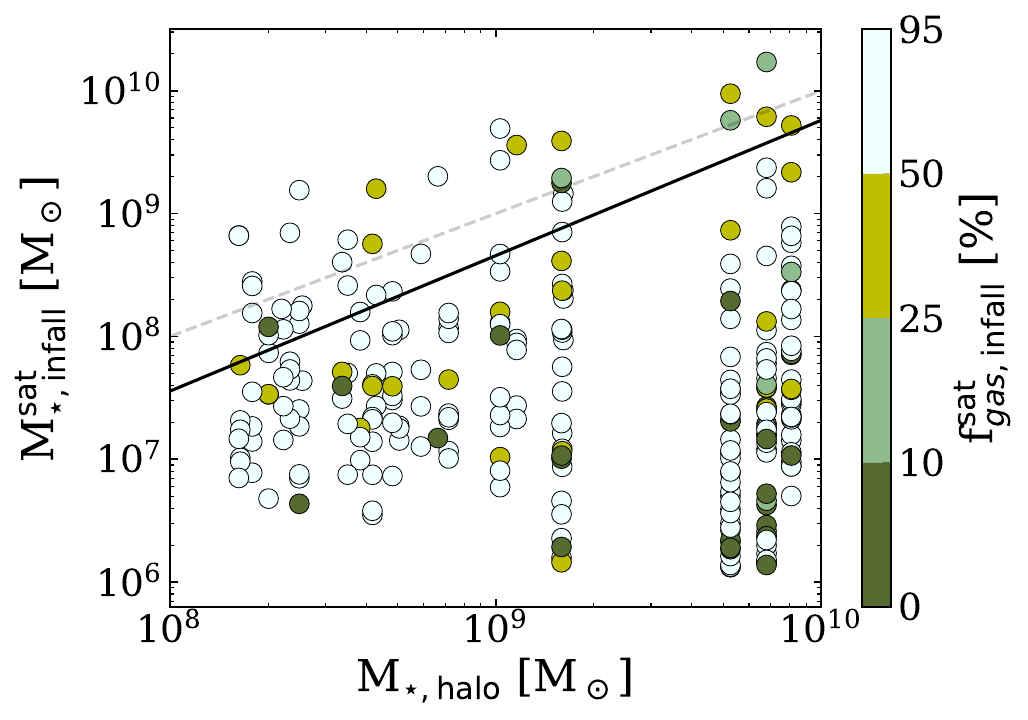}
    \caption{Upper panel: Stellar mass of the two main satellite contributors. Lower panel: All the accreted satellites with M$^{\rm sat}_{\star, \rm infall}>10^7\Msun$. Both panels are color-coded by the satellite gas fraction at infall time. The dashed line represents the 1:1 line, while the solid line indicates the linear regression.} 
    \label{fig:sats_gasfrac}
\end{figure}

Figure~\ref{fig:sats_gasfrac} illustrates the gas fraction of satellites contributing to the formation of stellar halos, defined as f$_{\rm gas, \rm infall}^{\rm sat} = \rm M_{\rm gas, \rm infall}^{\rm sat} / \rm  M_{\rm baryonic, \rm infall}^{\rm sat}$ . The upper panel shows the two SHMCs, while the lower panel displays all accreted satellites. Nearly all SHMCs in low-mass stellar halos have gas fractions exceeding $50\%$, whereas those in high-mass halos tend to exhibit lower gas fractions.

High-mass stellar halos primarily receive gas contributions from less massive satellites that fell in at higher redshifts, suggesting their endo-debris stars are preferentially formed in these satellites less massive than the two SHMCs. In contrast, low-mass stellar halos continue to receive gas contributions over time due to the fact that most of their infalling satellites have higher gas fractions. This ongoing gas accretion could also feed the star formation activity in the central galaxy, as is indicated by the ages of in-situ stars in Fig.~\ref{fig:SFH}.

\section{Chemical abundances of stellar populations}\label{sec:chemical_abund}

Since the stellar populations in the halos followed different formation channels and have a variety of assembly histories, we expect them to have different chemical abundances \citep{Tissera_2012, Deason_2016, Monachesi_2016a,Harmsen_2017}. Satellites contributing to the formation of stellar halos of different masses display diverse properties, and hence we expect differences in the chemistry of the stellar halo populations. In the next section, we investigate whether these variations in chemical abundances might be used as indicators of the different formation channels.

\subsection{The mass-metallicity relation of the stellar halos}\label{ssec:mzr}

Observational and theoretical evidence shows that the stellar halos of Milky Way mass-sized galaxies exhibit a strong correlation between their metallicity and stellar mass \citep[MZhR;][]{Deason_2016, Harmsen_2017, Bell_2017, DSouza_2018}. In this section, we examine the presence of the MZhR in our sample.

Figure~\ref{fig:FeH_Pop_z0} shows the MZhR for our stellar halos (black pentagons) and the corresponding three defined stellar populations, in-situ (blue circles), endo-debris (light green circles), and ex-situ (orange circles). There is a clear correlation between [Fe/H] and $\mhalo$, indicated by the Spearman correlation factors. Each population shows a different MZhR, which can be described by the following linear regression:
\begin{align}
&\rm{[Fe/H]_{ex-situ}} = 0.27 \times \rm{log}_{10}(\mhalo) - 3.70, \nonumber\\
&\rm{[Fe/H]_{endo-debris}}= 0.25 \times \rm{log}_{10}(\mhalo) - 3.27, \nonumber\\
&\rm{[Fe/H]_{in-situ}} = 0.29 \times \rm{log}_{10}(\mhalo) - 3.46, \\
&\rm{[Fe/H]_{halo}} = 0.23 \times \rm{log}_{10}(\mhalo) - 3.46. \nonumber
\end{align}

The in-situ population in Fig.~\ref{fig:FeH_Pop_z0} is split in the two formation paths, disk-heated (blue line) and clumpy (dashed blue line) stars, described in \ref{sec:pop}. These two populations exhibit different ages and metallicities. In-situ disk heated stars show higher [Fe/H] and a steeper MZhR than the endo-debris and ex-situ components. The combination of the different populations drives the global stellar halo MZhR.

The existence of the MZhR stems from the fact that the stellar halos assembled principally for accreted material from satellite galaxies, and galaxies followed the MZR. Hence, we expect that their combination would reflect a similar behavior \citep{Tissera_2014, Deason_2016, Bell_2017, DSouza_2018, Monachesi_2019} as it is the case for the \cielo~halos. Regarding the in-situ populations, most massive galaxies have more enriched ISM as they follow the MZR, which allows them to contribute with higher enriched disk-heated stars as is shown by our results.
Hence, consistent with previous findings, we confirm that SHMCs play a key role in driving this relation. However, at a given $\mhalo$ there is a variety of assembly histories, as is shown in Fig.~\ref{fig:sats_caract} and Fig.~\ref{fig:sats_gasfrac}, which could contribute to the dispersion of the MZhR (Fig.~\ref{fig:FeH_Pop_z0}). This aligns with the results of \citet{DSouza_2018}, which attributed the dispersion to the diversity of accretion histories involved in forming the accreted stellar halo at a given mass. In an upcoming article, we shall conduct a detailed analysis of the MZhR and its evolution with redshift.

\subsection{[O/Fe] - [Fe/H] plane}\label{sec:OFe_FeH}

In this section, we also investigate the [O/Fe]-[Fe/H] plane, which has been used to study the origin of stars in the Galactic components due to the interplay between the chemical production of Oxygen and Iron. These chemical elements are produced mainly by different supernova types with different lifetimes. The delayed enrichment by SNIa relative to SNII causes [O/Fe] to decrease as [Fe/H] increases, leading to the appearance of a knee \citep{Tinsley_79}. Nearby disk galaxies \citep{Sharma_2021} and our Milky Way \citep{Haywood_2013} show at least two different sequences leading a knee in the $\alpha-$plane in the disk component, which has been attributed to different formation channels (thin and thick disk). However, radial stellar migration, fresh gas infall, and the accretion of stars from different systems can affect the chemical abundances of the ISM and the star-formation activity in each system, which might introduce significant variations in the characteristics of the $\alpha-$Fe knee. In what follows, we investigate the presence of the $\alpha-$Fe knee in our stellar halos.

\begin{figure}
    \centering
    \includegraphics[width=0.45\textwidth]{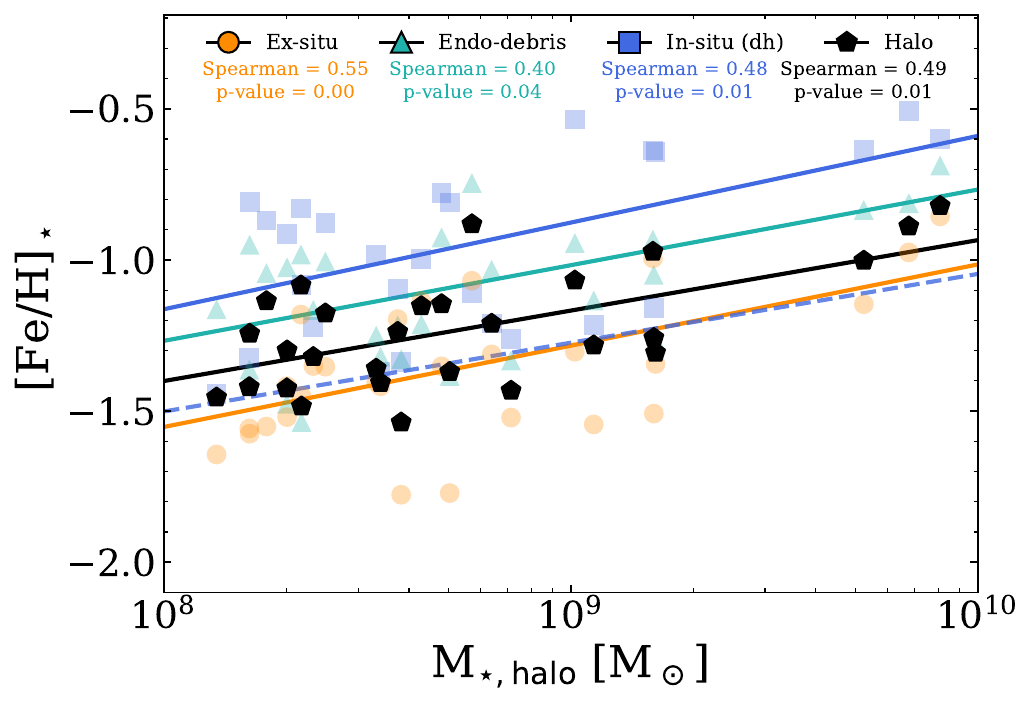}
    \caption{Stellar halo MZR (black pentagons). We also display the corresponding relations for ex-situ stars (orange circles), endo-debris stars (light green triangles), and in-situ disk-heated stars (blue squares). We fit linear regressions to the data for each population (solid colored lines codes as the corresponding symbols). Additionally, we added the linear fit for the clumpy in-situ stars (dashed blue).}
    \label{fig:FeH_Pop_z0}
\end{figure}

\begin{figure}
    \centering
    \includegraphics[width=0.48\textwidth]{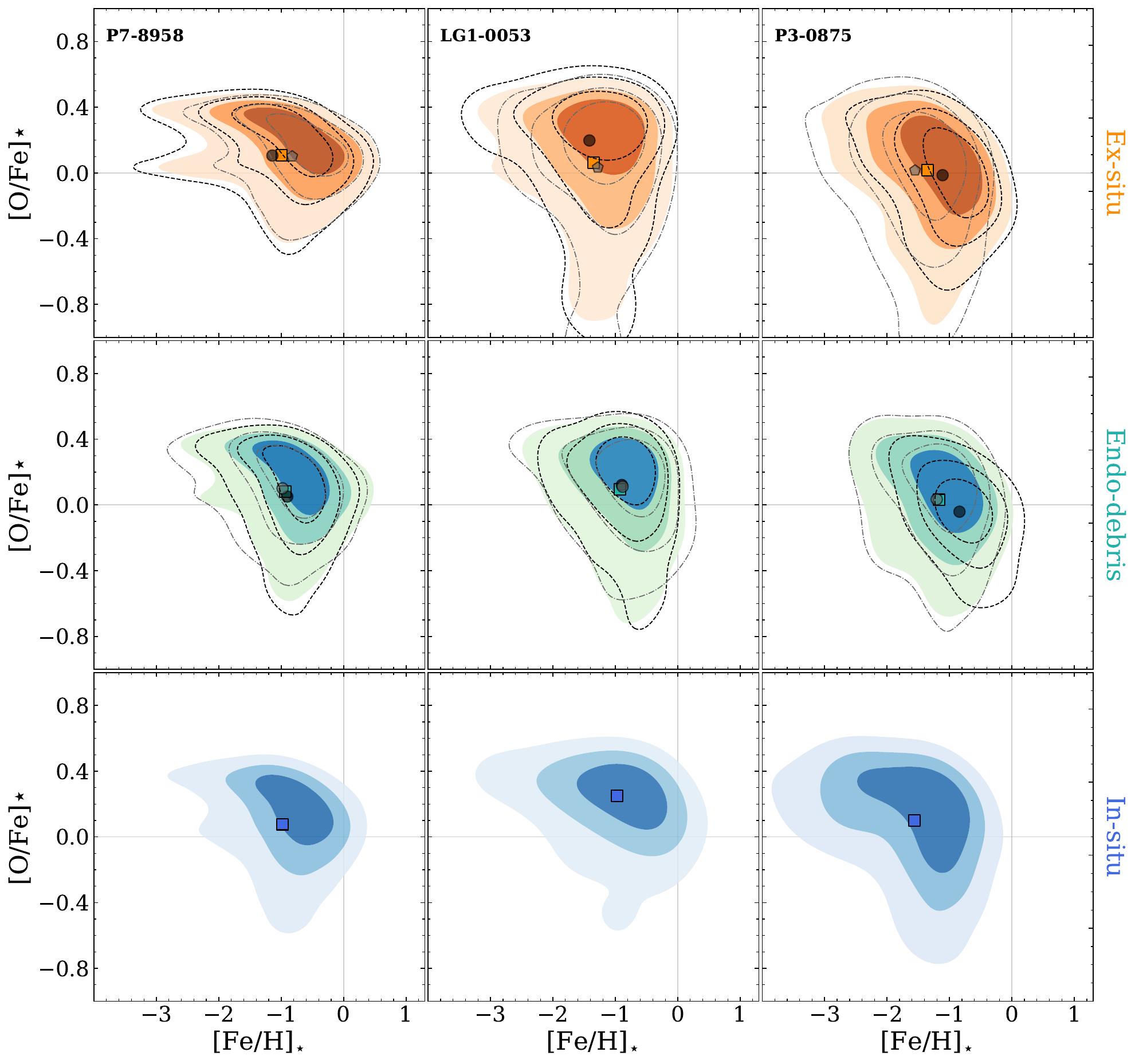} 
    \caption{[O/Fe]-[Fe/H] plane for the stellar particles of each population in the stellar halo of three typical galaxies taken as examples: ex-situ (orange), endo-debris (light green), and in-situ (blue). Median values are shown by squares. Contours represent the $25\%$, $50\%$, and $75\%$ of the peak density.
Black(gray) contours show the stars contributed by the first main contributor SHMC1 (black) and the second one, SHMC2 (gray).}.
    \label{fig:OFe_FeH_example}
\end{figure}

We selected three typical galaxies with different galaxy/stellar halo masses and different accretion histories to illustrate the distribution of the chemical abundances in the simulated stellar halos. Figure~\ref{fig:OFe_FeH_example} shows the distribution of [O/Fe] and [Fe/H] for the three populations studied (ex-situ, endo-debris, and in-situ) in the three galaxies. P7-8958 (left panel) has a massive stellar halo ($\mhalo = 1.59\times 10^9 \Msun$), LG1-0053 (middle panel) has an intermediate-mass halo ($\mhalo = 4.82\times 10^8 \Msun$) and P3-0875 (right panel) has a low-mass stellar halo ($\mhalo = 2.33\times 10^8 \Msun$). The isocontours that enclose the $25\%$, $50\%$, and $75\%$ of the mass are shown. The dotted contours show the stars contributed by the two SHMCs, SHMC1 (dashed-black line) and SHMC2 (dashed-dot gray line). The median value of the chemical abundances is shown as a square (stellar halo) or circle (SHMCs). Detailed values of each chemical abundance, masses of stellar halos, and their SHMCs are summarized in Table~\ref{table:abundances} for these three galaxies.

Figure~\ref{fig:OFe_FeH_example} illustrates the diversity in the formation histories of stellar halos and the chemical enrichment process of the structures that build them up. P7-8958 (left panel) clearly shows the expected knee in the [O/Fe]-[Fe/H] plane of the three stellar halo populations. Based purely on the distribution of its stellar populations on the $\alpha-$plane we would expect it to build up from massive systems, which are more metal-rich. On the other hand, LG1-0053 (middle panel) and P3-0875 (right panel) show a slightly different enrichment history that leads to the absence of the knee for the ex-situ and endo-debris populations, suggesting an early quenched star formation before the onset of SNIa. \cite{Horta_2022} also found an absence of the $\alpha-\rm{Fe}$ knee in the distribution of halo-accreted stars in two halo substructures of our Milky Way, Heracles and Thamnos. In particular, galaxies LG1-0053 (middle panel) and P3-0875 (right panel) have intermediate and low-mass stellar halos that exhibit lower [O/Fe] abundances, showing a tail extended to lower [O/Fe] in the ex-situ distributions. These galaxies accreted low-mass systems, $\minfall < 10^9 \Msun$, in agreement with the expected low-$\alpha$ and metal-poor stars found in dwarf satellite galaxies \cite{Tolstoy_2009}. These accreted stars provide metal-poor content that moves to lower values in the median of the [Fe/H] and [O/Fe] distribution of ex-situ stars.

In-situ stellar populations displayed in the third row of Fig.~\ref{fig:OFe_FeH_example} show a clear knee $\alpha-$plane for the three cases. This bimodality might be related to the two channels of formation mentioned in section \ref{sec:pop}. Since disk-heated and clumpy in-situ stars exhibit a difference of $\sim 3\rm Gyr$, and therefore there was enough time for SNIa to enrich the ISM from which they formed. On the other hand, endo-debris populations (second row of Fig.~\ref{fig:OFe_FeH_example}) show an anticorrelation that is stronger for P7-8958 but it is also present in the other two examples. We expect this clearer trend because these stars tend to be younger than ex-situ stars (Fig.~\ref{fig:SFH}) and hence, there has been more time for SNIa to contribute substantially to the enrichment of the ISM.

\begin{figure*}
    \centering
    \includegraphics[width=1\textwidth]{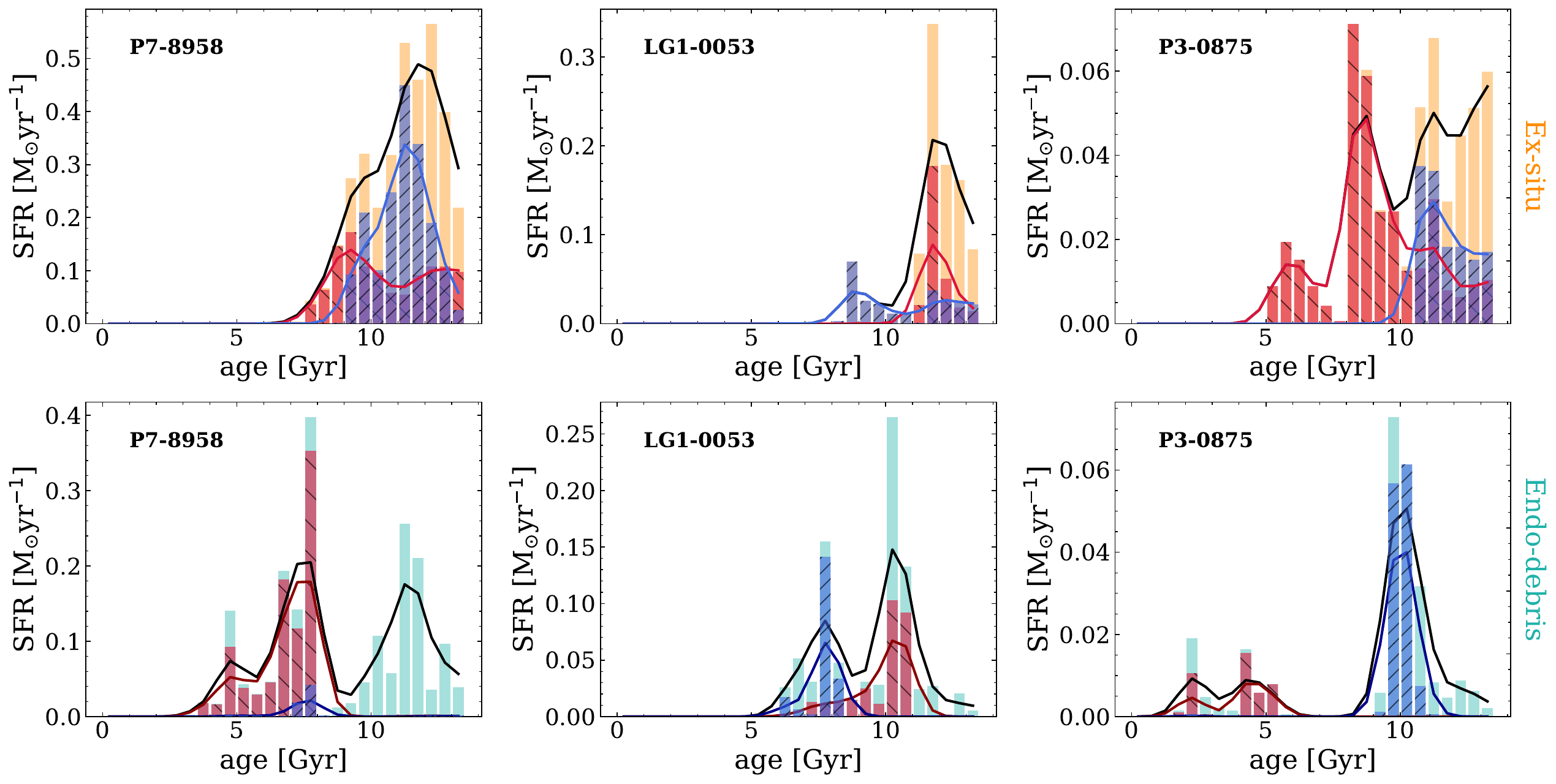} 
    \caption{Star formation history of the ex-situ (upper panel) and endo-debris (lower panel) stars in the stellar halos (orange and light green histograms, respectively, and solid black lines) and the corresponding distributions for stars contributed by the SHMC1 (red histograms and solid lines) and the SHMC2 (blue histograms and lines). }
    \label{fig:SFH_example}
\end{figure*}

In order to further investigate the origin of these trends, we studied the SFH of each accreted population and the specific stellar content contributed by the two SHMCs. We displayed the information for the same aforementioned three galaxies in Fig.~\ref{fig:SFH_example} to illustrate the different formation histories and their link to the different accretion events.
The upper panels of Fig.~\ref{fig:SFH_example} display the SFHs of the total ex-situ stars and the corresponding contributions of the ex-situ stars by the SHMC1 and the SHMC2.
The lower panel shows similar distributions for the endo-debris stars (light green).

The contribution to the ex-situ stars from the two SHMCs follows the peaks of the global distribution (orange histogram). In the case of the endo-debris (lower panels), for the most massive galaxy (left panel), the contribution of two SHMCs does not account for the whole star formation history, implying that they are not the only source of endo-debris stars and they contribute mostly to the younger endo-debris population. This picture agrees with Fig.~\ref{fig:sats_caract} upper panel and Fig.~\ref{fig:sats_gasfrac}, where we showed that for high-mass halos, the SHMCs contributed no more than 50 percent to the stellar halo with lower gas fractions. The rest of the accreted stars come from smaller systems with high gas fractions that could contribute to endo-debris populations, reflecting that the SHMCs for high-mass stellar halos are not the main contributors of endo-debris stars, and therefore will not necessarily determine the chemical abundance of the populations.

The SHMC1 of LG1-0053 provides $\alpha$-rich stars that came mainly from a single birth at high redshift (second-panel Fig.~\ref{fig:SFH_example}), while the SHMC2 provides $\alpha$-low stars due to an extended star formation in that satellite (see values detailed in Table~\ref{table:abundances}. Both satellite galaxies contribute with the $\sim 60 \%$ of the accreted stellar halo. The assembly of the remnant stellar halo comes mainly from low-mass satellite galaxies ($\msatmc < 10^8 \mstar$) accreted at z>1.5. They contribute with metal-poor material ([Fe/H]$\sim$-1.3) that is less $\alpha$ enhanced, i.e median [O/Fe]$\sim$ 0.02 and percentiles range [$25^{\rm th}$,$75^{\rm th}$] = [0.02,0.07].
The SHMC1 of galaxy P3-0875 exhibits extended SFH giving time to SNIa to enrich the medium and decrease the [O/Fe] abundances.
On the other hand, the two SHMCs of P7-8958 build the $75\%$ of the accreted stellar halo mass. They are massive satellite galaxies ($\msatmc> 10^8 \mstar$) accreted at z$\sim 1$. As can be seen in Fig.~\ref{fig:OFe_FeH_example} they contribute with a tail at lower metallicities; however, the median value of ex-situ stars is more metal-rich than low-mass stellar halos (Fig.~\ref{fig:FeH_Pop_z0}).

\begin{figure}
    \centering
    \includegraphics[width=0.45\textwidth]{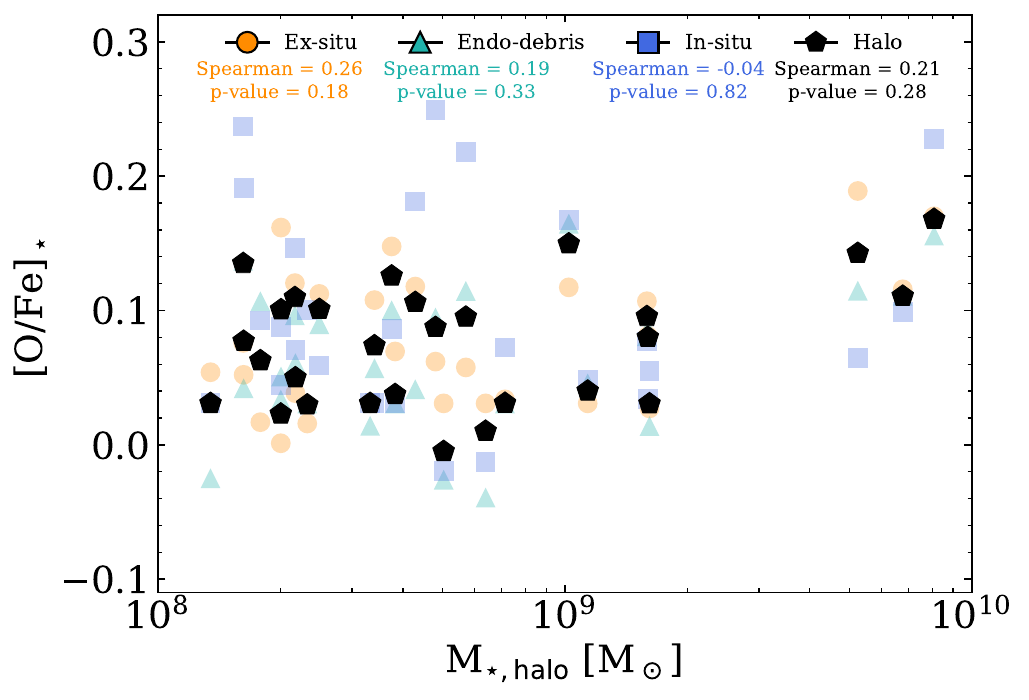}
    \caption{Median chemical abundances of [O/Fe] for the stellar halo and its stellar populations as a function of the stellar halo mass (see color and symbols code in Fig.~\ref{fig:FeH_Pop_z0}). Spearman coefficient and p-values are presented.}
    \label{fig:OFe_Mhalo}
\end{figure}

\begin{table*}[]
\caption{Properties of three CIELO galaxies selected for illustration purposes.}
\resizebox{\textwidth}{!}{%
\begin{tabular}{cccccccccccc}
\hline
Sim & ID gal & $\mhalo$ & M$_{\rm{SHMC1}}$ & M$_{\rm{SHMC2}}$ & Pop
     & [Fe/H]$_{\rm{\star,halo}}$ & [Fe/H]$_{\rm{SHMC1}}$ & [Fe/H]$_{\rm{SHMC2}}$  
     & [O/Fe]$_{\rm{\star,halo}}$ & [O/Fe]$_{\rm{SHMC1}}$ & [O/Fe]$_{\rm{SHMC2}}$ \\
     & & {[}$\Msun${]} & {[}$\Msun${]} & {[}$\Msun${]} &
     & {[}dex{]} & {[}dex{]} & {[}dex{]} & {[}dex{]} & {[}dex{]} & {[}dex{]} \\ \hline
     & & & & & Ex-situ & -0.99 & -1.14 & -0.82 & 0.10 & 0.11 & 0.10 \\
P7 & 8958 & 1.59e+09 & 3.90e+09 & 1.77e+09
     & Endo-debris & -0.93 & -0.90 & -0.98 & 0.08 & 0.05 & 0.10 \\
     & & & & & In-situ & -0.97 & & & 0.07 & & \\ \hline
     & & & & & Ex-situ & -1.35 & -1.41 & -1.28 & 0.06 & 0.19 & 0.03 \\
LG1 & 0053 & 4.80e+08 & 1.02e+08 & 2.33e+08
    & Endo-debris & -0.93 & -0.89 & -0.88 & 0.09 & 0.12 & 0.11 \\
    & & & & & In-situ & -0.97 & & & 0.24 & & \\ \hline
    & & & & & Ex-situ & -1.35 & -1.11 & -1.55 & 0.01 & -0.01 & 0.01\\
P3 & 0875 & 2.33e+08 & 6.97e+08 & 6.22e+07
    & Endo-debris & -1.16 & -0.83 & -1.20 & 0.03 & -0.04 & 0.03 \\
    & & & & & In-situ & -1.55 & & & 0.10 & & \\ \hline
\end{tabular}}
    \tablefoot{From left to right: simulation code; identification number galaxies; $\mhalo$, the stellar mass ranging from $1.5\ropt$ to $\rvir$; M$_{\rm{SHMC1}}$ and M$_{\rm{SHMC2}}$, the stellar masses at infall time of the satellite main satellites contributors; the median [Fe/H] and [O/Fe] for the stellar halo ([Fe/H]$_{\star,\rm{halo}}$, [O/Fe]$_{\star,\rm{halo}}$); and the median abundances ([Fe/H]$_{\rm{SHMC1}}$, [Fe/H]$_{\rm{SHMC2}}$ and ([O/Fe]$_{\rm{SHMC1}}$, [O/Fe]$_{\rm{SHMC2}}$) of the stellar populations brought by the two main satellite contributors. }
    \label{table:abundances}
\end{table*}

Considering that the stellar populations exhibit [O/Fe]-[Fe/H] characteristics, we now explore if the overallmedian [O/Fe] of the stellar population in the halo correlates with its stellar mass. Figure~\ref{fig:OFe_Mhalo} shows the [O/Fe] for the stellar halo and each of its stellar populations as a function of $\mhalo$. \cielo~stellar halos exhibit a flat trend for $\mhalo \leq 10^{9}\rm \Msun$ with the median [O/Fe] distributed within a range of $~0.15$ dex. This indicates no correlation between the level of $\alpha-$enrichment and the stellar halo masses. We quantify this by estimating the Spearman correlation factors which are shown in the figure. However, high-mass stellar halos have higher levels of [O/Fe]. Considering the low number of galaxies with high-mass stellar halos in our sample, this trend has to be taken with caution, and more statistics are needed to draw a robust conclusion. Nevertheless, this suggests a more complex combination of stellar populations with a diverse history of formation as we show above, which is responsible for the larger spread of median [O/Fe].

In fact, the most massive stellar halos have higher levels of [O/Fe] consistent with have been formed from more massive satellites which contributed with higher $\alpha$-enriched stars.
The higher $\alpha$-Fe is achieved because more massive systems tend to have stronger starbursts at higher redshift, forming a more significant fraction of $\alpha$-enriched stars. Hence, the low stellar mass halo, which receives lower mass satellites, also has a larger contribution of lower [O/Fe] stellar populations when the satellites have been able to form stars for an extended period at a lower rate \citep{Matteucci_1990} or had several starbursts.
However, it is also important to consider the particular characteristics of each star formation history to assess the relative importance of the high and low $\alpha$-enriched populations formed in a given satellite as is shown in Fig.~\ref{fig:SFH_example}.

\begin{figure}
    \centering
    \includegraphics[width=0.45\textwidth]{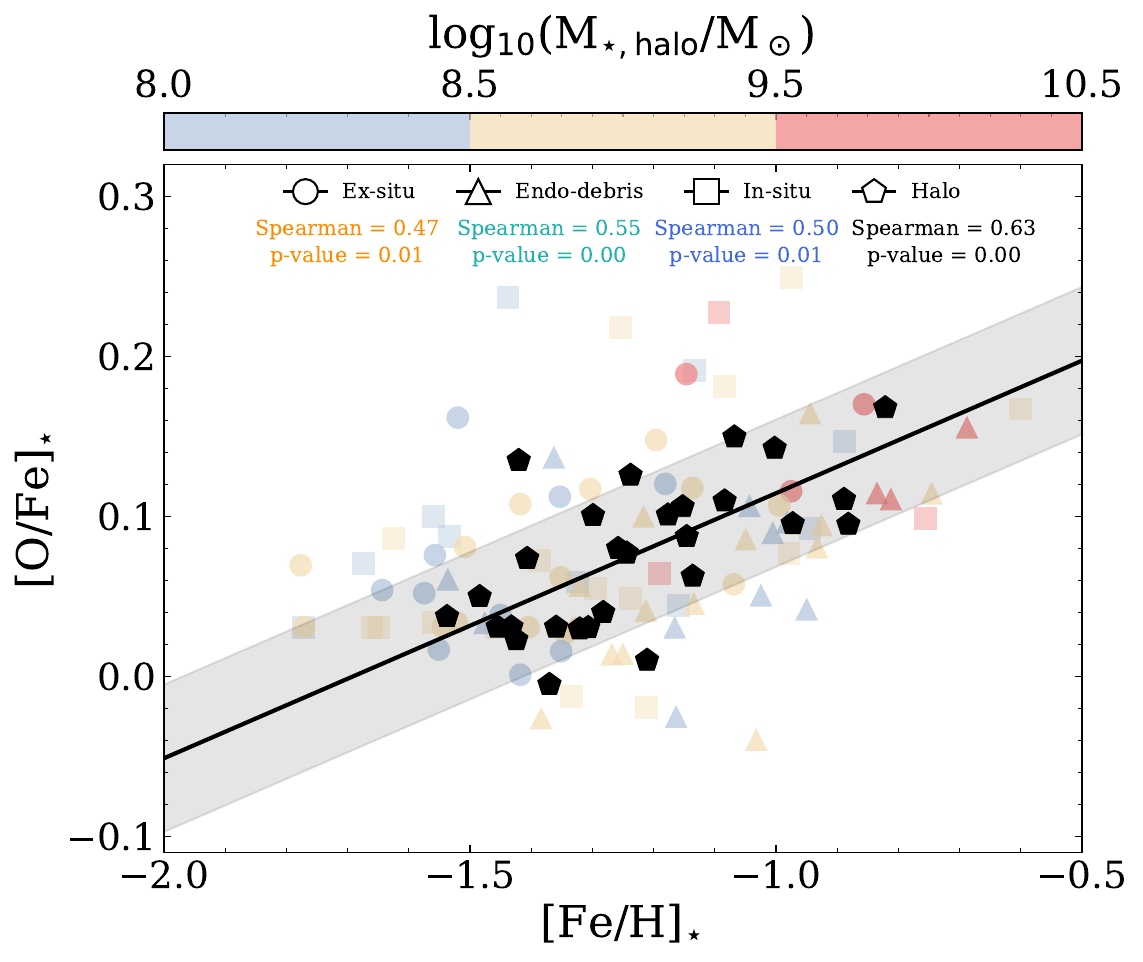}
    \caption{Median [O/Fe] and [Fe/H] for the stellar halo and each of its stellar populations: in-situ, endo-debris, and ex-situ, color-coded by the stellar halo mass (see color and symbols code in Fig.~\ref{fig:FeH_Pop_z0}). The linear fit for the relation for the whole stellar halo abundances is included (black line; the shadow region is delimited by standard deviation).}
    \label{fig:OFe_FeH_colorM}
\end{figure}

Figure~\ref{fig:OFe_FeH_colorM} displays the medians of [O/Fe] versus [Fe/H] per galaxy for the whole stellar halo and the three defined stellar populations. In this case, interestingly, the Spearman correlation factors are significant, indicating a correlation for all cases. Hence, we fitted a linear regression in the [O/Fe]-[Fe/H] plane, which can be described as follows,
\begin{align}
&\rm{[O/Fe]_{ex-situ}} = 0.11\times [Fe/H]_{\rm halo} + 0.22, \nonumber\\
&\rm{[O/Fe]_{endo-debris}}= 0.15\times [Fe/H]_{\rm halo} + 0.23, \nonumber\\
&\rm{[O/Fe]_{in-situ}} = 0.08\times [Fe/H]_{\rm halo} + 0.17, \\
&\rm{[O/Fe]_{halo}} = 0.16\times [Fe/H]_{\rm halo} + 0.28. \nonumber
\end{align}

The relation determined by endo-debris is steeper than those of ex-situ stars, mainly because of the lower levels of [O/Fe] in low-mass halos. This can be understood considering that they have formed from smaller satellites which also tend to lower star formation activity and be more gas-rich. We expect these stars to reach lower [O/Fe] for lower [Fe/H] \citep{Matteucci_1990}. We also estimated the stellar mass fraction with
[O/Fe]$< 0.1$~dex contributed by the SMHCs. We found that the first contributor is the one that brings in, on average, 50 percent, but with a large variety (within a range of 20-90 percent). Therefore, for some halos, contributions from the rest of the accreted systems, smaller than the SMHCs, have a significant impact. The global $[\alpha$/Fe]-[Fe/H] relation is a product of the combinations of stellar populations from all satellites and hence, stores more information on the assembly history. We shall further explore further this relation in a forthcoming paper.

\section{Conclusions}\label{secc:summary}
We analyzed 28 galaxies with stellar masses in the range of $[10^9, 10^{11}]\Msun$, selected from zoom-in simulations, to investigate the assembly history of stellar halos. The stellar halos were identified using the AM-E method \citep{Tissera_2012}, focusing on the outer region, spanning from $1.5\ropt$ to $\rvir$. Three distinct stellar populations— in-situ, endo-debris, and ex-situ stars—were classified based on their formation channels. These populations are differentiated by their median chemical abundances and ages, reflecting their unique formation histories.
Our main conclusions are summarized as follows:

   \begin{enumerate}
\item Stellar halos are mostly made of accreted material, a combination of ex-situ stars and endo-debris stars. However, the mass fraction of each population does not correlate with the stellar halo mass. The mass distribution as a function of radius varies with the systems, but the mass fraction of endo-debris stars is generally higher at lower radii, while the ex-situ stars dominate the outskirts. \\

      \item The three stellar populations identified follow different formation channels which are reflected in their chemical abundances and ages. The median [Fe/H] of each population correlates with the stellar halo mass. In fact, the metallicity of ex-situ and endo-debris stars drives the stellar halo MZR observed, due to their contribution to the bulk of stellar mass to the halo.\\ 

      \item In-situ stars are the youngest population and they are affected by the gas accretion from satellite galaxies, which is the fuel for star formation in the host galaxy. Their SFR will be linked to their accretion history. Low-mass stellar halos exhibit younger in-situ stars than high-mass stellar halos. On the other hand, ex-situ stars are the oldest population. \\ 
      
      \item There is a correlation between stellar halo mass and the stellar mass of the main contributor. More massive stellar halos tend to accrete more massive satellites, which following the MZR, are more metal-enriched. This supports previous works and the correlation observed in Fig.~\ref{fig:FeH_Pop_z0}, where the metallicity of each population increases with the stellar halo mass. \\
      
      \item The number of satellites contributing 90 percent of the accreted stellar halo mass varies from three (25th percentile) to eight (75th percentile) with a median of four. Stellar halos more massive than $\mhalo>10^{9.5}\Msun$ need, on a median, 5 more satellites to build up the $90\%$ of its stellar halo mass than low mass stellar halos, $\mhalo<10^{8.5}\Msun$.\\

    \item The distribution of alpha abundances of the stellar populations allows us to build a general picture of the chemical evolution of the stellar halos. Differences in the star formation history of the contributing satellites will reveal differences in the $\alpha$-enrichment levels. The stellar populations of the \cielo~ halos show the expected trend on the [O/Fe]-[Fe/H], which globally agree with observational evidence, when considering Milky Way-mass galaxies \citep{Naidu_2020, Horta_2022}. However, since stellar halos are a collection of stars from different satellite galaxies, this diversity in star formation histories leads to a weak correlation between the median [O/Fe] and the stellar halo masses, at least for $\mhalo \leq 10^{9}\rm \Msun$.
      \\

      \item 
    We found a correlation between the median [O/Fe] and the median [Fe/H] of the stellar halos. 
    This trend can be understood considering that stellar halos are forged by the mix of contributing from low and high $\alpha$-enriched stars from satellite galaxies of different masses \citep{Matteucci_1990}. More massive stellar halos exhibit higher [O/Fe] levels, consistent with contributions from more massive satellites that underwent intense starbursts at higher redshifts, producing a larger fraction of $\alpha$-enriched stars and reaching higher levels of [Fe/H]. In contrast, low-mass stellar halos, primarily formed from smaller satellites, include a relatively greater contribution of lower [O/Fe] populations. These populations originate from satellites with prolonged or intermittent low star formation histories, which hence can reach lower [O/Fe].

   \end{enumerate}

CIELO galaxies are well suited for understanding and interpreting the diversity of stellar halo properties in the nearby Universe and other environments. The variety of accretion and merger histories allows us to study and characterize correlations and scaling relations between stellar halo properties and the formation history of their host galaxies. We have found not only well-defined MZhR but also a clear [$\alpha$/Fe]-[Fe/H] relation that stores more comprehensive information of the assembly history of halos. Here, we only explored the outer regions of the stellar halos. The inner halo might have different relative contributions from stellar populations formed by different formation channels, along with its coexistence with the disk component, both of which will be analyzed in a forthcoming paper.

Finally, observations of stellar halos have increased in the last few years and will continue to do so in the coming with current and future datasets such as the GALAH \citep{Buder_2022} and 4MOST \citep{Survey_4most} surveys. They will provide a wide chemical space in which to identify different enrichment histories, and hence contribute to unveiling the assembly paths of galaxies and their stellar halos.

\begin{acknowledgements}
    We thank the anonymous referee for his/her comments which help improving the clarity of this manuscript. We also thank the useful discussion with people from the EvolGal4D team:  Isha Shailesh, Anell Cornejo, Catalina Casanueva, Benjamin Silva, Doris Stoppacher, Valentina Miranda, Francisco Jara, Ignacio Muñoz, Daniela Barrientos, Maria Luiza L. Dantas.
      JGJ acknowledges funding by ANID (Beca Doctorado Nacional, Folio 21210846) as well as the support from the Núcleo Milenio ERIS. 
      PBT acknowledges partial funding by Fondecyt-ANID 1240465/2024 and Núcleo Milenio ERIS NCN2021\_017. This project has received funding from the European Union Horizon 2020 Research and Innovation Programme under the Marie Sklodowska-Curie grant agreement No 734374- LACEGAL. We acknowledge partial support by ANID BASAL project FB210003.
AM gratefully acknowledges support by FONDECYT Regular grant 1212046 as well as funding from the Max Planck Society through a ``PartnerGroup” grant.
ES acknowledges funding by Fondecyt-ANID Postdoctoral 2024 Project N°3240644 and thanks the Nucleo Milenio ERIS.
DP gratefully acknowledges financial support from ANID through FONDECYT Postdoctrorado Project 3230379. DP also acknowledges support from the ANID BASAL project FB210003. BTC gratefully acknowledges funding by ANID (Beca Doctorado Nacional, Folio 21232155).
              
This project made use of the Ladgerda Cluster (Fondecyt 1200703/2020 hosted at the Institute for Astrophysics, Chile), the NLHPC (Centro de Modelamiento Matem\'atico, Chile) and the Barcelona Supercomputer Center (Spain).
\end{acknowledgements}

%
%
\bibliographystyle{aa} 
\bibliography{bibliography.bib}

\end{document}